\documentclass[mathematics,article,accept,moreauthors,12pt,a4paper]{mdpi} % if eps figure are used, use this line with LaTeX and dvi2pdf %only
\lastpage{28}
\doinum{10.3390/math2010012}
\pubvolume{2}
\pubyear{2014}
\history{Received: 10 October 2013; in revised form: 26 January 2014 / Accepted:  26 January 2014 / \\
Published: 14 February 2014}

\usepackage{soul}
\usepackage{amssymb,amsmath}
\usepackage{graphicx}
\usepackage{subfigure}
\makeatletter
\renewcommand{\@thesubfigure}{\normalsize(\textbf{\alph{subfigure}})}
\makeatother
\usepackage{array, dcolumn, multirow}
\usepackage{hyphenat}
\usepackage{enumitem,caption,booktabs}
\setitemize{parsep=0pt,itemsep=0pt,leftmargin=.4in}
\setenumerate{parsep=0pt,itemsep=0pt,leftmargin=.4in}

\usepackage[sort&compress]{natbib}

\Title{On the Folded Normal Distribution}

\Author{Michail Tsagris $^{1}$, Christina Beneki $^{2}$ and Hossein Hassani $^{3,}$*}

\address{%
$^{1}$ School of Mathematical Sciences, University of Nottingham, NG7 2RD, UK; \linebreak E-Mail: mtsagris@yahoo.gr \\
$^{2}$ School of Business and Economics, TEI of Ionian Islands, 31100 Lefkada, Greece; \linebreak E-Mail: christinabeneki@gmail.com \\
$^{3}$ Statistical Research Centre, Executive Business Centre, Bournemouth University,  BH8 8EB,  UK; }

\corres{E-Mail: hhassani@bournemouth.ac.uk; \linebreak Tel.: +44-120-296-8708.}

\abstract{The characteristic function of the folded normal
distribution and its moment function are derived.
 The entropy of the folded normal distribution and the Kullback--Leibler from the normal
 and half normal distributions are approximated using Taylor series.
 The accuracy of the results are also assessed using different criteria.
 The maximum likelihood estimates and confidence intervals for the parameters
 are obtained using the asymptotic theory and bootstrap method. The coverage of the confidence intervals is also examined.}


\begin{document}
%%%%%%%%%%%%%%%%%%%%%%%%%%%%%%%%%%%%%%%%%%%%%%%%%%%%%%%%%%%%
\vspace{-12pt}
% main text
\section{Introduction}
\label{Sec1}

Mainly studied in the 1960s, the folded normal distribution is a special case of the Gaussian distribution occurring when the sign of the variable is always positive. In 1961, a method of estimating the parameters based upon the estimating equations of the moments was discussed in \cite{leone1961}, where they also gave some examples of its applications in the industrial sector. The folded normal distribution was used to study the magnitude of deviation of an automobile strut alignment \cite{Lin}. The properties of the multivariate folded normal distribution with its possible applications were studied in \cite{Ashis}.
In addition, tables with probabilities for a range of values of the vector of parameters were provided, and an application of the model with real data was illustrated. An alternative method using the second and fourth moments of the distribution was proposed in \cite{elandt1961}, whilst \cite{johnson1962} performed maximum likelihood estimation and calculated the asymptotic information matrix. Thereafter, the sequential probability ratio test for the null hypothesis of the location parameter being zero against a specific alternative was evaluated in \cite{johnson1963} with the idea of illustrating the use of cumulative sum control charts for multiple observations.

In \cite{sundberg1974}, the author dealt with the hypothesis testing of the zero location parameter regardless of the variance being known or not. The distribution formed by the ratio of two folded normal variables was studied and illustrated with a few applications in \cite{KK}. The folded normal distribution has been applied to many practical problems. For instance, introduced in \cite{liao2010} is an economic model to determine the process specification limits for folded normally distributed data.

%The folded normal distribution is a particular type of distribution which however does not incorporate the nice properties of the normal distribution. For example it is not a stable distribution, (the distribution of the sum is not a folded normal distribution) and it does not belong to the exponential family of distributions. In addition, there is no sufficient statistic for the two parameters and the maximum likelihood estimates are accessible only via numerical optimization. However, it has been applied to many practical problems.

Through this paper, we will examine the folded normal distribution from a different perspective. In the process, we will consider the study of some of its properties, namely the characteristic and moment generating functions, the Laplace and Fourier transformations and the mean residual life of this distribution. The entropy of this distribution and its Kullback--Leibler divergence from the normal and half normal distributions will be approximated via the Taylor series. The accuracy of the approximations are assessed using numerical examples.

Also reviewed here is the maximum likelihood estimates (for an introduction, see \cite{leone1961}), with examples from simulated data given for illustration purposes. Simulation studies will be performed to assess the validity of the estimates with and without bootstrap calibration in low sample cases. Numerical optimization of the log-likelihood will be carried out using the simplex method \citep{nelder1965}.

\section{The Folded Normal}

\begin{figure}[H]
\centering
\begin{tabular}{cc}
\includegraphics[scale=0.5]{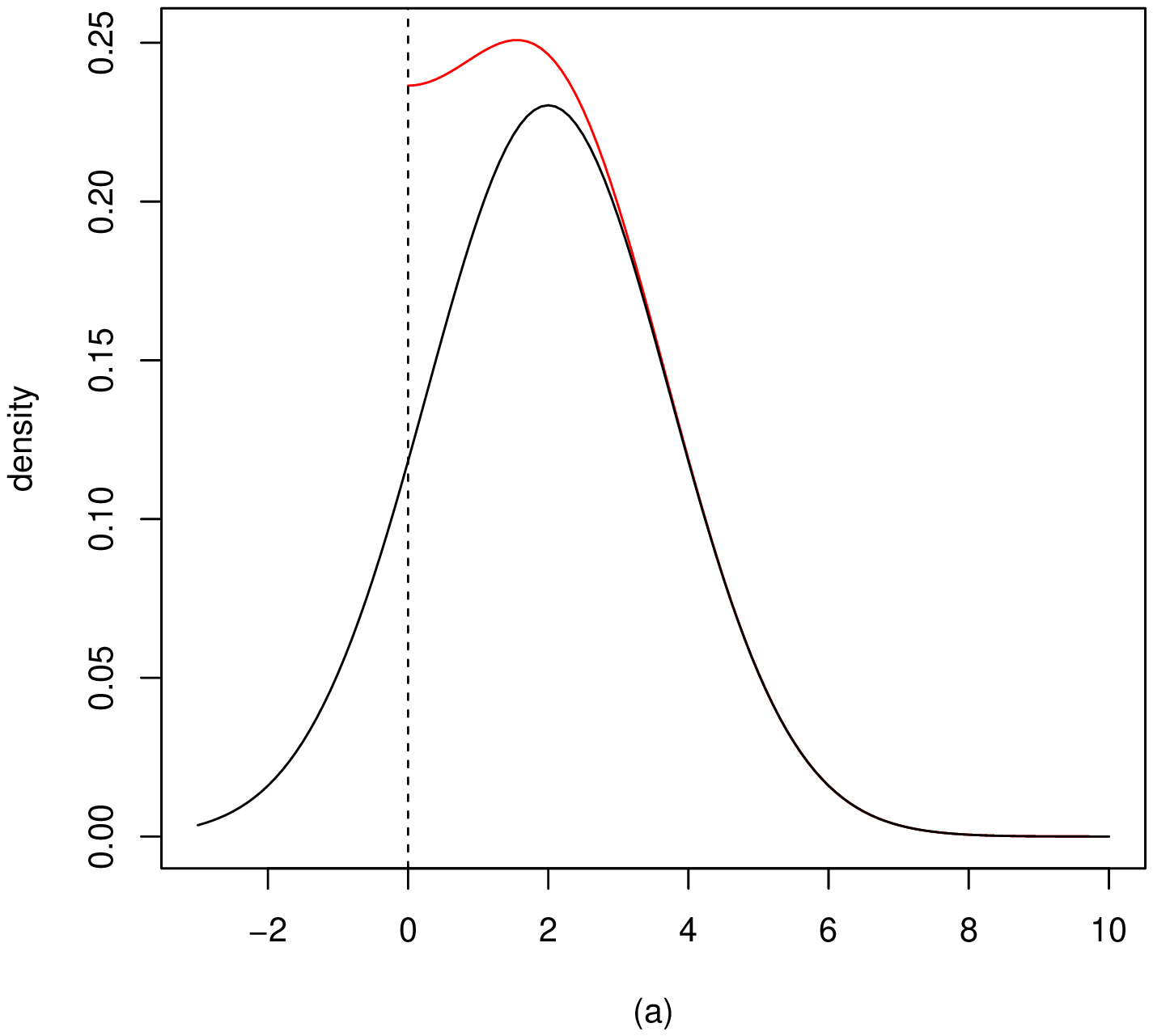} &
\includegraphics[scale=0.5]{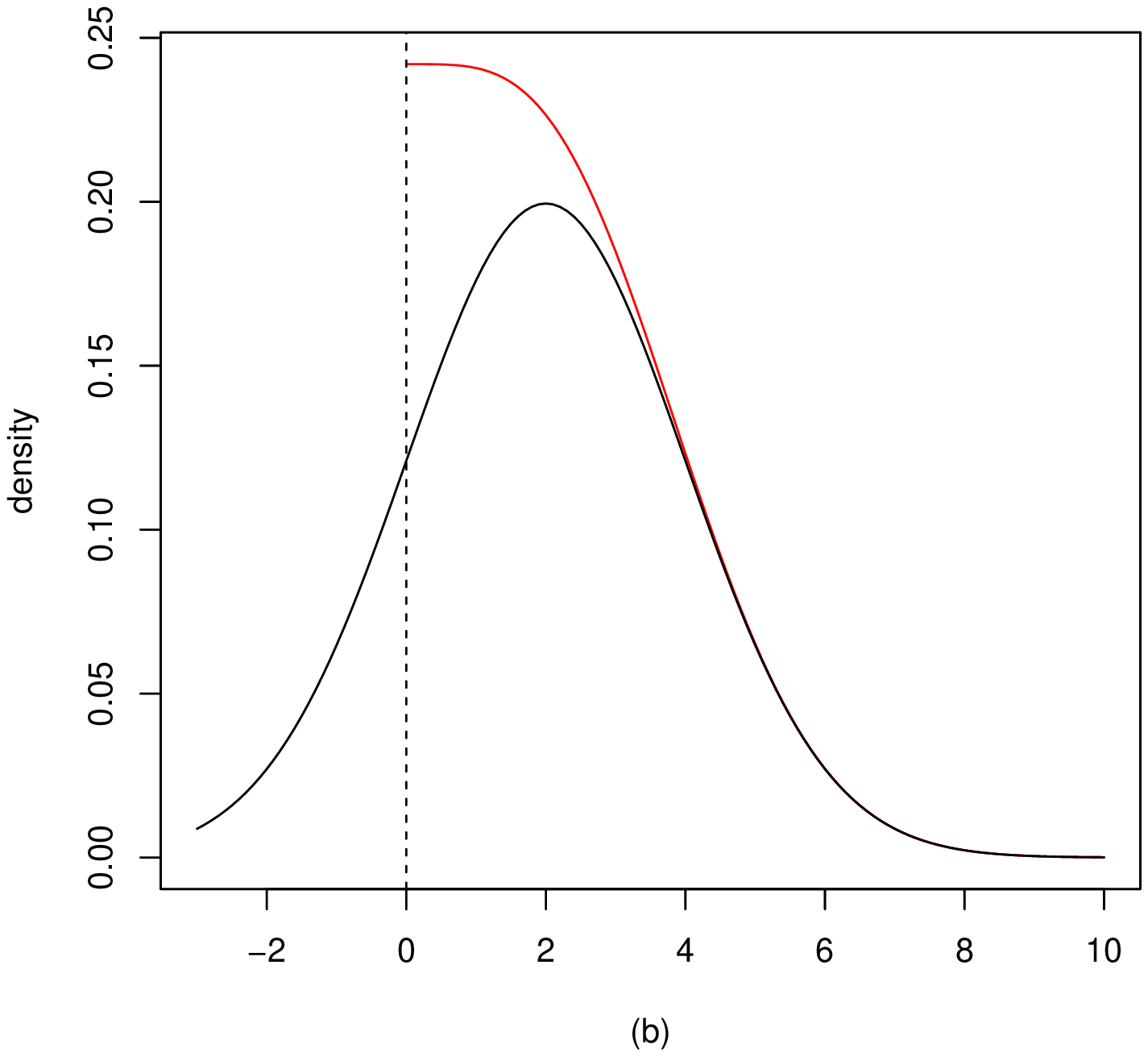}
\end{tabular}
\caption{The black line is the density of the $N\left(\mu,\sigma^2\right)$ and the red line of the $FN\left(\mu,\sigma^2\right)$. The parameters in the left figure ({\bf a}) are $\mu=2$ and $\sigma^2=3$ and in the right figure  ({\bf b})  $\mu=2$ and $\sigma^2=4$.}
\label{density}
\end{figure}

The folded normal distribution with parameters $\left(\mu,\sigma^2 \right)$ stems from taking the absolute value of a normal distribution with the same vector of parameters. The density of $Y$, with $Y$$\sim$$N\left(\mu, \sigma^2 \right)$ is \linebreak given by:
\begin{eqnarray}
f\left(y \right)=\frac{1}{\sqrt{2\pi\sigma^2}}e^{-\frac{1}{2\sigma^2}\left(y-\mu \right)^2}
\end{eqnarray}
Thus, $X=\left|Y \right|$, denoted by $Y\sim FN\left(\mu, \sigma^2 \right)$, has the following density:
\begin{eqnarray} \label{fd}
f\left(x \right)=\frac{1}{\sqrt{2\pi\sigma^2}}\left[ e^{-\frac{1}{2\sigma^2}\left(x-\mu \right)^2}+
e^{-\frac{1}{2\sigma^2}\left(x+\mu \right)^2} \right]
\end{eqnarray}
The density can be written in a more attractive form \cite{johnson1962}:
\begin{eqnarray}
f\left(x \right)=\sqrt{\frac{2}{\pi\sigma^2}}e^{-\frac{\left(x^2+\mu^2 \right)^2}{2\sigma^2}}\cosh{\left(\frac{\mu x}{\sigma^2}\right)}
\end{eqnarray}
and by expanding the $cosh$ via a Taylor series, we can also write the density as:
\begin{eqnarray}
f\left(x \right)=\sqrt{\frac{2}{\pi\sigma^2}}e^{-\frac{\left(x^2+\mu^2 \right)^2}{2\sigma^2}}\sum_{n=0}^\infty\frac{\left(-1\right)^n}{\left( 2n\right)!}\left(\frac{\mu x}{\sigma^2}\right)^{2n}
\end{eqnarray}
We can see that the folded normal distribution is not a member of the exponential family. The cumulative distribution can be written as:
\begin{eqnarray}
F\left(x \right)=\frac{1}{2}\left[erf\left(\frac{x-\mu}{\sqrt{2\sigma^2}}\right)+ erf\left(\frac{x+\mu}{\sqrt{2\sigma^2}}\right)\right]
\end{eqnarray}
where $erf$ is the error function:
\begin{eqnarray}
erf\left(x\right)=\frac{2}{\sqrt{pi}}\int_0^x e^{-t^2}dx
\end{eqnarray}
The mean and the variance of Equation (\ref{fd}) is calculated using direct calculation of the integrals\linebreak as follows \cite{leone1961}:
\begin{eqnarray} \label{mean}
\mu_f = \sqrt{\frac{2}{\pi}}\sigma e^{-\frac{\mu^2}{2\sigma^2}}+\mu\left[1-2\Phi\left(-\frac{\mu}{\sigma}\right) \right]
\end{eqnarray}
\begin{eqnarray} \label{var}
\sigma^2_f = \mu^2+\sigma^2-\mu_f^2
\end{eqnarray}
where $\Phi\left(.\right)$ is the cumulative distribution function of the standard normal distribution.
The third and fourth moments about the origin are calculated in \cite{elandt1961}.
We develop the calculation further by providing the characteristic function and the moment generating
function of Equation (\ref{fd}). Figure (\ref{density}) shows the densities of the folded normal for some parameter values.

\vspace{-18pt}

\subsection{Relations to Other Distributions} \label{relations}

The distribution of $Z=X/\sigma$ is\scalebox{.95}[1.0]{ a non-central} $\chi$ distribution with one degree of freedom and \linebreak non-centrality parameter equal to $(\mu/\sigma)^2$ \cite{johnson1994}. It is clear that when $\mu=0$, a central $\chi_1$ is obtained. The half normal distribution is a special case of Equation (\ref{fd}), with $\mu=0$ for which \cite{psarakis1990} showed that it is the limiting form of the folded (central) t distribution as the degrees of freedom of the latter go to infinity. Both distributions are further developed in the bivariate case in \cite{psarakis2000}.

The folded normal distribution can also be seen as the the limit of the folded non-standardized $t$ distribution as the degrees of freedom go to infinity. The folded non-standardized $t$ distribution is the distribution of the absolute value of the non-standardized $t$ distribution with $v$ degrees of freedom:
\begin{eqnarray}
g\left(x\right)=\frac{\Gamma\left(\frac{v+1}{2}\right)}{\Gamma\left(\frac{v}{2}\right)\sqrt{v\pi\sigma^2}}\left\lbrace
\left[1+\frac{1}{v}\frac{\left(x-\mu\right)^2}{\sigma^2}\right]^{-\frac{v+1}{2}}+\left[1+\frac{1}{v}\frac{\left(x+\mu\right)^2}{\sigma^2}\right]^{-\frac{v+1}{2}} \right\rbrace
\end{eqnarray}
\subsection{Mode of the Folded Normal Distribution}

The mode of the distribution is the value of $x$ for which the density is maximised. In order to \linebreak find this value, we take the first derivative of the density with respect to $x$ and set it equal to zero. Unfortunately, there is no closed form. We can, however, write the derivative in a better way and end up with a non-linear equation.
\begin{eqnarray}
\frac{df\left(x\right)}{dx}=0 & \Rightarrow & -\frac{\left(x-\mu\right)}{\sigma^2}e^{-\frac{1}{2}\frac{\left(x-\mu\right)^2}{\sigma^2}}-
\frac{\left(x+\mu\right)}{\sigma^2}e^{-\frac{1}{2}\frac{\left(x+\mu\right)^2}{\sigma^2}}=0 \\
& \Rightarrow & x\left[e^{-\frac{1}{2}\frac{\left(x-\mu\right)^2}{\sigma^2}}+e^{-\frac{1}{2}\frac{\left(x+\mu\right)^2}{\sigma^2}}\right]-
\mu \left[e^{-\frac{1}{2}\frac{\left(x-\mu\right)^2}{\sigma^2}}-e^{-\frac{1}{2}\frac{\left(x+\mu\right)^2}{\sigma^2}}\right]=0 \\
& \Rightarrow & x\left(1+e^{-\frac{2\mu x}{\sigma^2}}\right)-\mu\left(1-e^{-\frac{2\mu x}{\sigma^2}}\right)=0 \\
& \Rightarrow & \left(\mu+x\right)e^{-\frac{2\mu x}{\sigma^2}}=\mu-x \\
& \Rightarrow & x=-\frac{\sigma^2}{2\mu}\log{\frac{\mu-x}{\mu+x}}
\end{eqnarray}

We saw from numerical investigation that when $\mu<\sigma$, the maximum is met when $x=0$. When $\mu \geq \sigma$, the maximum is met at $x>0$, and when $\mu$ becomes greater than $3\sigma$, the \scalebox{.95}[1.0]{maximum approaches $\mu$.} This is of course something to be expected, since, in this case, the folded normal converges to the \linebreak normal distribution.

\subsection{Characteristic Function and Other Related Functions of the Folded Normal Distribution}

Forms for the higher moments of the distribution when the moment is an odd and even number is provided in \cite{elandt1961}. Here, we derive its characteristic and, thus, the moment generating function.
\begin{eqnarray} \label{phi}
\varphi_x\left(t\right)&=& E\left( e^{itX}\right)=\int_0^{\infty}e^{itx}f_X\left(x \right)dx =
\int_0^{\infty}e^{itx}\frac{1}{\sqrt{2\pi\sigma^2}}\left[ e^{-\frac{1}{2\sigma^2}\left(x-\mu \right)^2}+
e^{-\frac{1}{2\sigma^2}\left(x+\mu \right)^2} \right]dx \nonumber \\
&=& \int_0^{\infty}\frac{e^{itx-\frac{1}{2\sigma^2}\left(x-\mu \right)^2}}{\sqrt{2\pi\sigma^2}} dx+
\int_0^{\infty}\frac{e^{itx-\frac{1}{2\sigma^2}\left(x+\mu \right)^2}}{\sqrt{2\pi\sigma^2}}dx \nonumber \\
&=& \int_0^{\infty}\frac{e^A}{\sqrt{2\pi\sigma^2}}dx+\int_0^{\infty}\frac{e^B}{\sqrt{2\pi\sigma^2}}dx
\end{eqnarray}
We will work now with the forms $A$ and $B$.
\begin{eqnarray}
A&=& itx-\frac{1}{2\sigma^2}\left(x-\mu \right)^2=\frac{2i\sigma^2tx-x^2+2\mu x-\mu^2}{2\sigma^2}=
-\frac{x^2-2x\left(i\sigma^2 t+\mu \right)+\mu^2}{2\sigma^2} \\
&=& -\frac{\left[x-\left(i\sigma^2 t+\mu \right) \right]^2+\sigma^4 t^2-2i\sigma^2 t \mu}{2\sigma^2}=
-\frac{\left(x-a\right)^2}{2\sigma^2}-\frac{\sigma^2 t^2}{2}+i\mu t
\end{eqnarray}
where $ a=i\sigma^2 t+\mu $. Thus, the first part of Equation (\ref{phi}) becomes:
\begin{eqnarray}
\int_0^{\infty}\frac{e^A}{2\pi\sigma^2}dx &=& e^{\frac{-\sigma^2 t^2}{2}+i\mu t}\int_0^{\infty}\frac{e^{-\left(x-\alpha\right)^2}}{2\pi\sigma^2}dx=e^{\frac{-\sigma^2 t^2}{2}+i\mu t}\left[1-P\left(X \leq 0 \right) \right] \\
&=& e^{\frac{-\sigma^2 t^2}{2}+i\mu t}\left[1-\Phi\left(-\frac{a}{\sigma} \right) \right]=
e^{\frac{-\sigma^2 t^2}{2}+i\mu t}\left[1-\Phi\left(-\frac{\mu}{\sigma}-i\sigma t \right) \right]
\end{eqnarray}
The second exponent, $B$, using similar calculations becomes:
\begin{eqnarray}
B=itx-\frac{1}{2\sigma^2}\left(x+\mu \right)^2=-\frac{\left[x-\left(i\sigma^2 t -\mu \right) \right]^2}{2\sigma^2}-\frac{\sigma^2 t^2}{2}-i\mu t
\end{eqnarray}
and, thus, the second part of  Equation (\ref{phi}) becomes:
\begin{eqnarray}
\int_0^{\infty}\frac{e^B}{2\pi\sigma^2}dx=e^{-\frac{\sigma^2 t^2}{2}-i\mu t}\left[1-\Phi\left(\frac{\mu}{\sigma}-i\sigma t \right) \right]
\end{eqnarray}
Finally, the characteristic function becomes:
\begin{eqnarray} \label{charact}
\varphi_x\left(t\right)=e^{\frac{-\sigma^2 t^2}{2}+i\mu t}\left[1-\Phi\left(-\frac{\mu}{\sigma}+i\sigma t \right) \right]+
e^{-\frac{\sigma^2 t^2}{2}-i\mu t}\left[1-\Phi\left(\frac{\mu}{\sigma}+i\sigma t \right) \right]
\end{eqnarray}
Below, we list some more functions that include expectations.
\begin{enumerate}
\item The moment generating function of Equation (\ref{fd}) exists and is equal to:
\begin{eqnarray} \label{moment}
M_x\left(t\right)=\varphi_x\left(-it\right)=e^{\frac{\sigma^2 t^2}{2}+\mu t}\left[1-\Phi\left(-\frac{\mu}{\sigma}-\sigma t \right) \right]+
e^{\frac{\sigma^2 t^2}{2}-\mu t}\left[1-\Phi\left(\frac{\mu}{\sigma}-\sigma t \right) \right]
\end{eqnarray}
We can see that the characteristic generating function can be differentiated infinitely many times, since the first derivative contains the density of the normal distribution, and thus, it always contains some exponential terms. The folded normal distribution is not a stable distribution. That is, the distribution of the sum of its random variables do not form a folded normal distribution. We can see this from the characteristic (or the moment) generating function \scalebox{.95}[1.0]{Equation (\ref{charact}) or Equation (\ref{moment}).}
\vspace{0.5pt}
\item The cumulant generating function is simply the logarithm of the moment generating function:
\begin{eqnarray}
K_x\left(t\right)=\log{M_x\left(t\right)}=
\left(\frac{\sigma^2t^2}{2}+\mu t\right)\log{\left\lbrace 1-\Phi\left(-\frac{\mu}{\sigma}-\sigma t \right) +
e^{-2\mu t}\left[1-\Phi\left(\frac{\mu}{\sigma}-\sigma t \right) \right] \right\rbrace}
\end{eqnarray}
\item The Laplace transformation can easily be derived from the moment generating function and is equal to:
\begin{eqnarray}
E\left(e^{-tx}\right)=e^{\frac{\sigma^2t^2}{2}-\mu t}\left[1-\Phi\left(-\frac{\mu}{\sigma}+\sigma t \right) \right]+
e^{\frac{\sigma^2 t^2}{2}+\mu t}\left[1-\Phi\left(\frac{\mu}{\sigma}+\sigma t \right) \right]
\end{eqnarray}
\item The Fourier transformation is:
\begin{eqnarray} \label{fourier}
\hat{f}\left(t\right)=\int_{-\infty}^{\infty}e^{-2\pi ixt}f\left(x\right)dx=E\left(e^{-2\pi iXt}\right)
\end{eqnarray}
However, this is closely related to the characteristic function. We can see that \linebreak $E\left(e^{-2\pi ixt}\right)=\phi_x\left(-2\pi t\right)$. Thus, Equation
(\ref{fourier}) becomes:
\begin{eqnarray}
\hat{f}\left(t\right)=\phi_x\left(-2\pi t\right)= & & e^{\frac{-4\pi^2\sigma^2 t^2}{2}- i2\pi \mu t}\left[1-\Phi\left(-\frac{\mu}{\sigma}-i2\pi \sigma t \right) \right] \\
&+& e^{-\frac{4\pi^2 \sigma^2 t^2}{2}+i2\pi\mu t}\left[1-\Phi\left(\frac{\mu}{\sigma}-i2\pi \sigma t \right) \right]
\end{eqnarray}
\item The mean residual life is given by:
\begin{eqnarray}
E\left(X-t\vert X>t\right)=E\left(X\vert X>t\right)-t
\end{eqnarray}
where $t \in \mathbb{R}_+$. The above conditional expectation is given by:
\begin{eqnarray} \label{resid}
E\left(X\vert X>t\right)=\int_t^{\infty}\frac{xf\left(x\right)}{P\left(x>t\right)}dx=
\int_t^{\infty}\frac{xf\left(x\right)}{1-F\left(t\right)}dx
\end{eqnarray}
The denominator in Equation (\ref{resid}) is written as $1-\frac{1}{2}\left[erf\left(\frac{x-\mu}{\sqrt{2\sigma^2}}\right)+ erf\left(\frac{x+\mu}{\sqrt{2\sigma^2}}\right)\right]$. The contents within the integral in the numerator of Equation (\ref{resid}) could be replaced by $1-F\left(t\right)$, as well, but we will not replace it. The calculation of the numerator is done in the same way as the calculation of the mean. Thus:
\begin{eqnarray}
\int_t^{\infty}xf\left(x\right)dx &=& \int_t^{\infty}x\frac{1}{\sqrt{2\pi\sigma^2}}e^{-\frac{1}{2\sigma^2}\left(x-\mu \right)^2}dx+
\int_t^{\infty}x\frac{1}{\sqrt{2\pi\sigma^2}}e^{-\frac{1}{2\sigma^2}\left(x+\mu \right)^2} dx \\
&=& \frac{\sigma}{\sqrt{2\pi}}e^{\frac{\left(t-\mu\right)^2}{\sigma^2}}+\mu\left[1-\Phi\left(\frac{t-\mu}{\sigma}\right)\right]+
\frac{\sigma}{\sqrt{2\pi}}e^{\frac{\left(t-\mu\right)^2}{\sigma^2}}-\mu\Phi\left(\frac{t-\mu}{\sigma}\right) \\
&=& \sqrt{\frac{2}{\pi}}\sigma e^{\frac{\left(t-\mu\right)^2}{\sigma^2}}+\mu\left[1-2\Phi\left(\frac{t-\mu}{\sigma}\right)\right]
\end{eqnarray}
\noindent Finally, Equation (\ref{resid}) can be written as:
\begin{eqnarray}
E\left(X-t\vert X>t\right)=\frac{\sqrt{\frac{2}{\pi}}\sigma e^{\frac{\left(t-\mu\right)^2}{\sigma^2}}+\mu\left[1-2\Phi\left(\frac{t-\mu}{\sigma}\right)\right]}{1-\frac{1}{2}\left[erf\left(\frac{x-\mu}{\sqrt{2\sigma^2}}\right)+ erf\left(\frac{x+\mu}{\sqrt{2\sigma^2}}\right)\right]}-t
\end{eqnarray}
\end{enumerate}

\section{Entropy and Kullback--Leibler Divergence}

When studying a distribution, the entropy and the Kullback--Leibler divergence from some other distributions are two measures that have to be calculated. In this case, we tried to approximate both of these quantities using a Taylor series. Numerical examples are displayed to show the performance of \linebreak the approximations.

\subsection{Entropy}

The entropy is defined as the negative expectation of $-\log{f\left(x\right)}$.
\begin{eqnarray} \label{entro}
E &=& E\left[-\log{f\left(x\right)}\right]=-\int_0^\infty\log{f\left(x\right)}f\left(x\right)dx \nonumber \\
&=& -\int_0^\infty f\left(x\right)\log{\left\lbrace\frac{1}{\sqrt{2\pi\sigma^2}}\left[ e^{-\frac{1}{2\sigma^2}\left(x-\mu \right)^2}+
 e^{-\frac{1}{2\sigma^2}\left(x+\mu \right)^2} \right]\right\rbrace}dx \nonumber \\
&=& \log{\sqrt{2\pi\sigma^2}}\int_0^\infty f\left(x\right)dx
-\int_0^\infty f\left(x\right)\log{\left[e^{-\frac{\left(x-\mu\right)^2}{2\sigma^2}}\left(1+\frac{e^\frac{\left(x+\mu\right)^2}{2\sigma^2}}{e^{-\frac{\left(x-\mu\right)^2}{2\sigma^2}}} \right) \right]} dx \nonumber \\
&=& \log{\sqrt{2\pi\sigma^2}}+\int_0^\infty \frac{x^2-2\mu x+\mu^2}{2\sigma^2}f\left(x\right)-
\int_0^\infty f\left(x\right)\log{\left(1+e^{-\frac{2\mu x}{\sigma^2}}\right)} dx
\end{eqnarray}

Let us now take the second term of Equation (\ref{entro}) and see what is equal to:

\begin{eqnarray}
\frac{1}{2\sigma^2}\int_0^\infty x^2f\left(x\right)=\frac{\mu^2+\sigma^2}{2\sigma^2} \ \ \text{by exploiting the knowledge of variance Equation (\ref{var})}
\end{eqnarray}
\begin{eqnarray}
\frac{-2\mu}{2\sigma^2}\int_0^\infty xf\left(x\right)=-\mu \frac{\mu_f}{\sigma^2} \ \ \text{since the first moment is given in Equation (\ref{mean}) and}
\end{eqnarray}
\begin{eqnarray}
\frac{\mu^2}{2\sigma^2}\int_0^\infty f\left(x\right)=\frac{\mu^2}{2\sigma^2}
\end{eqnarray}
Finally, the third term of Equation (\ref{entro}) is equal to:
\begin{eqnarray}
A_n=-\int_0^\infty f\left(x\right)\sum_{n=1}^\infty \frac{\left(-1\right)^{n+1}}{n}e^{-\frac{2n\mu x}{\sigma^2}} dx
\end{eqnarray}
by making use of the Taylor expansion for $\log{\left(1+x\right)}$ around zero, but instead of $x$, we have $e^{-\frac{2\mu x}{\sigma^2}}$.
Thus, we have managed to ``break'' the second integral of entropy Equation (\ref{entro}) down to smaller \linebreak pieces of:
\begin{eqnarray*}
A_n &=&-\int_0^\infty \sum_{n=1}^\infty \frac{\left(-1\right)^{n+1}}{n}e^{a_nx}\frac{1}{\sqrt{2\pi\sigma^2}}e^{-\frac{1}{2\sigma^2}\left(x-\mu \right)^2}dx-\int_0^\infty \sum_{n=1}^\infty \frac{\left(-1\right)^{n+1}}{n}e^{a_nx}\frac{1}{\sqrt{2\pi\sigma^2}}e^{-\frac{1}{2\sigma^2}\left(x+\mu \right)^2}dx \\
&=& -\sum_{n=1}^\infty \frac{\left(-1\right)^{n+1}}{n}e^{\frac{\left(\mu+a_n\sigma^2\right)^2-\mu^2}{2\sigma^2}}
\left[1-\Phi\left(-\frac{\mu}{\sigma}-\frac{a_n}{\sigma} \right) \right] \\
& & - \sum_{n=1}^\infty \frac{\left(-1\right)^{n+1}}{n}e^{\frac{\left(\mu-a_n\sigma^2\right)^2-\mu^2}{2\sigma^2}}
\left[1-\Phi\left(\frac{\mu}{\sigma}-\frac{a_n}{\sigma} \right) \right]
\end{eqnarray*}
by interchanging the order of the summation and the integration, filling up the square in the same way to the characteristic function and with $a_n=-\frac{2n\mu}{\sigma^2}$. The final form of the entropy is given in Equation (\ref{entrofinal}):
\begin{eqnarray} \label{entrofinal}
E & \simeq & \log{\sqrt{2\pi\sigma^2}}+\frac{1}{2}+\frac{\mu^2-\mu \mu_f}{\sigma^2}-\sum_{n=1}^\infty \frac{\left(-1\right)^{n+1}}{n}e^{\frac{\left(\mu-2n\mu\right)^2-\mu^2}{2\sigma^2}}
\left[1-\Phi\left(-\frac{\mu}{\sigma}+\frac{-\frac{2n\mu}{\sigma^2}}{\sigma} \right) \right] \nonumber \\
& & - \sum_{n=1}^\infty \frac{\left(-1\right)^{n+1}}{n}e^{\frac{\left(\mu-2n\mu\right)^2-\mu^2}{2\sigma^2}}
\left[1-\Phi\left(\frac{\mu}{\sigma}+\frac{-\frac{2n\mu}{\sigma^2}}{\sigma} \right) \right]
\end{eqnarray}

\begin{figure}[H]
\centering
\begin{tabular}{cc}
\includegraphics[scale=0.5]{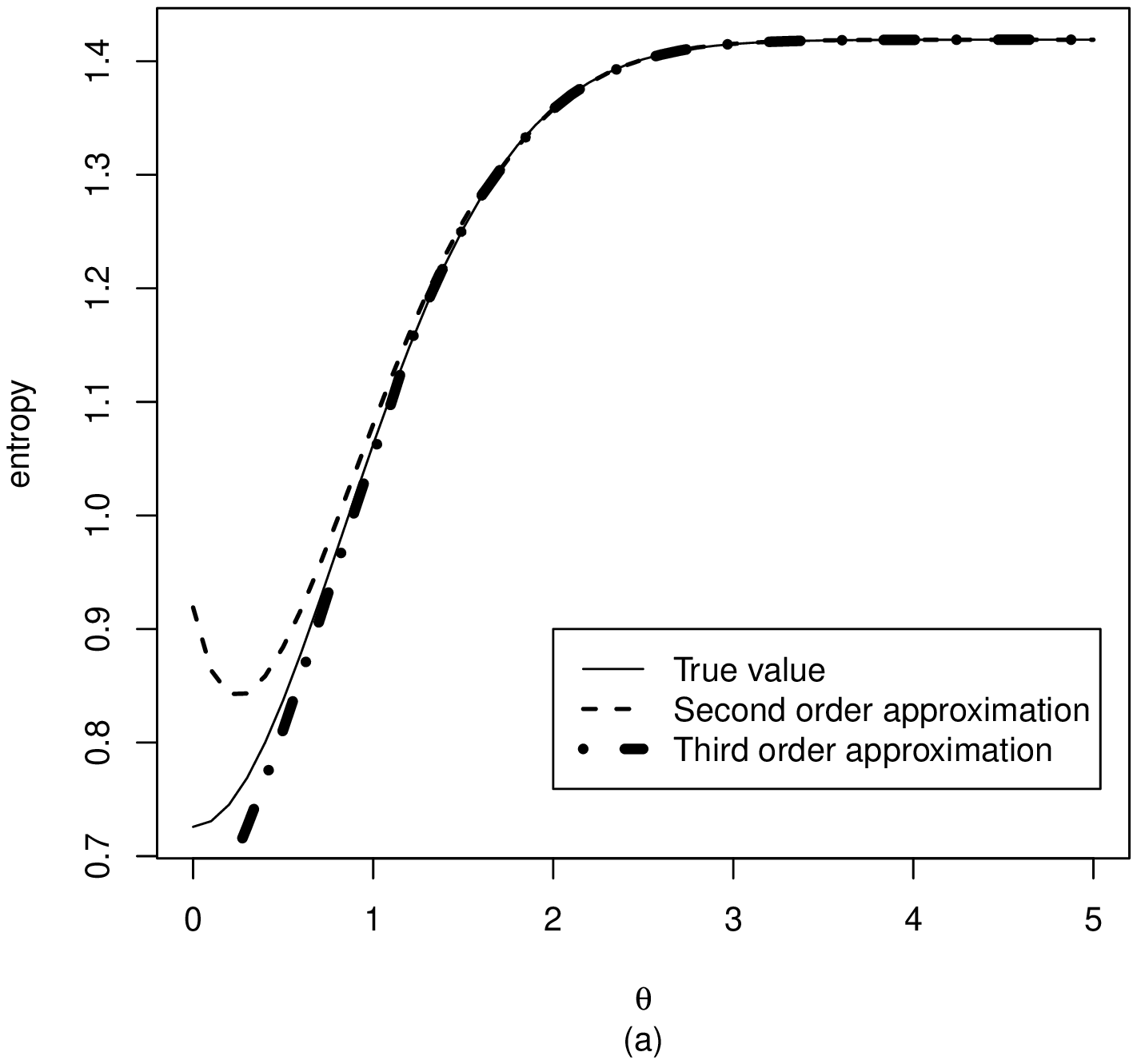} &
\includegraphics[scale=0.5]{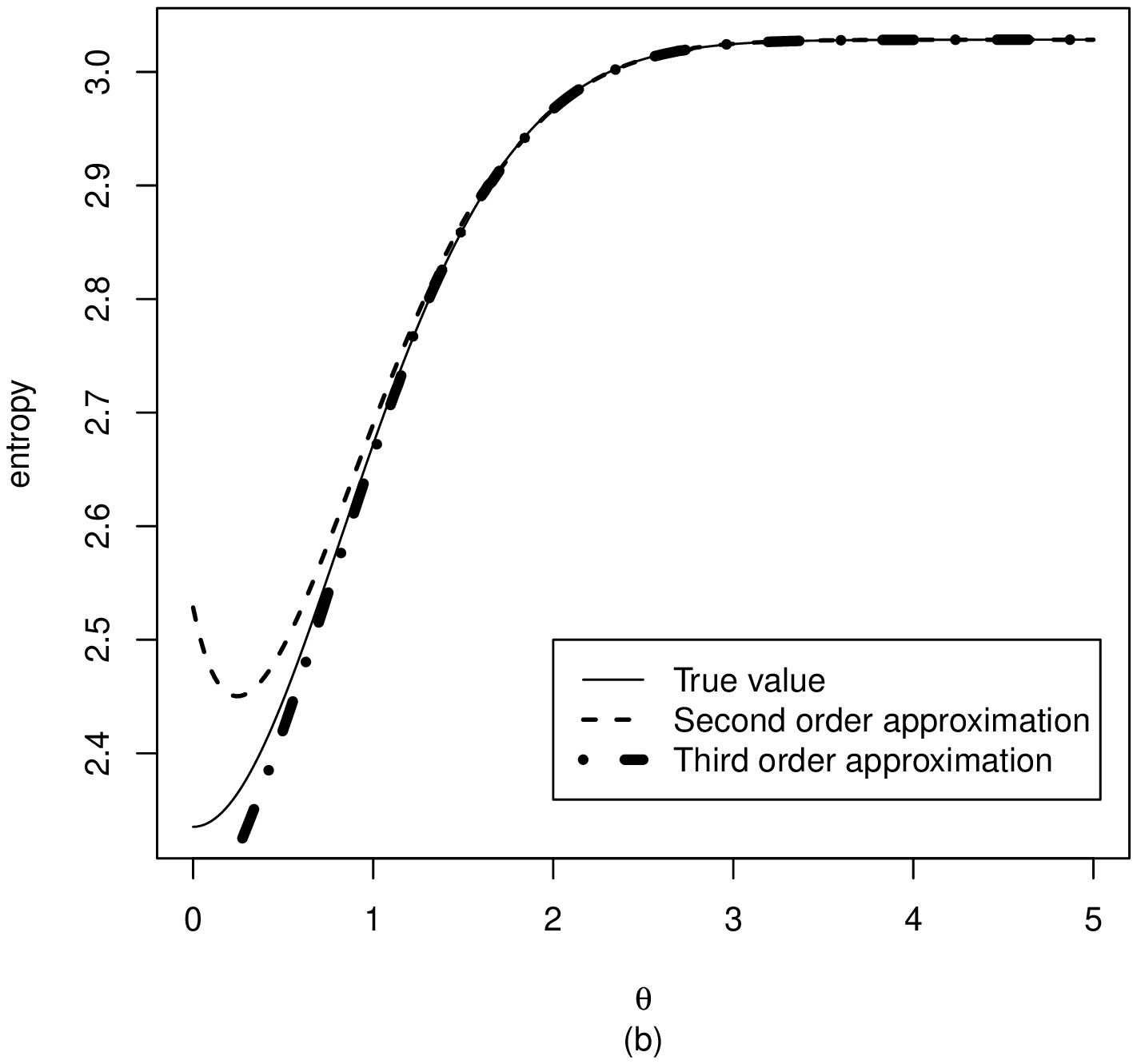}
\end{tabular}
\caption{Entropy values for a range of values of $\theta=\frac{\mu}{\sigma}$ with $\sigma=1$ ({\bf a}) and $\sigma=5$ ({\bf b}).}
\label{entrograph}
\end{figure}

Figure \ref{entrograph} shows the true value of Equation (\ref{entrofinal}), when $\sigma=5$ and $\mu$ ranges from zero to $25$, thus for values of $\theta=\frac{\mu}{\sigma}$ from zero to five. The true value was calculated using numerical integration. R%please define
 provides this option with the command \textit{integrate}.
The second and third order approximations (using the first two and three terms of the infinite sums in Equation (\ref{entrofinal})), are also displayed for comparison.

We can see that the second order approximation is not as good as the third order, especially for small values of $\theta$. The Taylor approximation of Equation (\ref{entrofinal}) is valid when the value, $a_n$, is close to zero. As with the logarithm approximation, the expansion is around zero; thus, when we start going further away from zero, the approximation loses its accuracy. The same is true in our case. When the values of $\theta$ are small, then the value of $\log{\left(1+e^{-\frac{2\mu x}{\sigma^2}}\right)}$ is far from zero. As $\theta$ increases, and, thus, the exponential term decreases, the Taylor series approximates true value better.
This is why we see a small discrepancy of the approximations on the left of Figure \ref{entrograph}, which become negligible later on.

\subsection{Kullback--Leibler Divergence from the Normal Distribution}

The Kullback--Leibler divergence \cite{Kullback1997} of one distribution from another in general is defined as the expectation of the logarithm of the ratio of the two distributions with respect to the first one:
\begin{eqnarray*}
KL\left(f \vert g\right)=E_f\left[\log{\frac{f}{g}} \right]=\int f\left(x\right)\log{\frac{f\left(x\right)}{g\left(x\right)}}dx
\end{eqnarray*}
The divergence of the folded normal distribution from the normal distribution is equal to:
\begin{eqnarray*}
KL(FN \vert \vert N) &=& \int_0^\infty\frac{1}{\sqrt{2\pi\sigma^2}}\left[ e^{-\frac{1}{2\sigma^2}\left(x-\mu \right)^2}+
e^{-\frac{1}{2\sigma^2}\left(x+\mu \right)^2} \right]\log{\frac{\frac{1}{\sqrt{2\pi\sigma^2}}\left[ e^{-\frac{1}{2\sigma^2}\left(x-\mu \right)^2}+
e^{-\frac{1}{2\sigma^2}\left(x+\mu \right)^2} \right]}{\frac{1}{\sqrt{2\pi\sigma^2}}e^{-\frac{1}{2\sigma^2}\left(x-\mu \right)^2}}} dx \\
&=& \int_0^\infty\frac{1}{\sqrt{2\pi\sigma^2}}\left[ e^{-\frac{1}{2\sigma^2}\left(x-\mu \right)^2}+
e^{-\frac{1}{2\sigma^2}\left(x+\mu \right)^2} \right]\log{\left(1+e^{-\frac{2\mu x}{\sigma^2}}\right)} dx
\end{eqnarray*}
which is the same as the second integral of Equation (\ref{entro}). Thus, we can approximate this divergence by the same Taylor series:
\begin{eqnarray*}
KL(FN \vert \vert N) & \simeq & \sum_{n=1}^\infty \frac{\left(-1\right)^{n+1}}{n}e^{\frac{\left(\mu-2n\mu\right)^2-\mu^2}{2\sigma^2}}
\left[1-\Phi\left(-\frac{\mu}{\sigma}+\frac{-\frac{2n\mu}{\sigma^2}}{\sigma} \right) \right] \nonumber \\
& & + \sum_{n=1}^\infty \frac{\left(-1\right)^{n+1}}{n}e^{\frac{\left(\mu-2n\mu\right)^2-\mu^2}{2\sigma^2}}
\left[1-\Phi\left(\frac{\mu}{\sigma}+\frac{-\frac{2n\mu}{\sigma^2}}{\sigma} \right) \right]
\end{eqnarray*}

\begin{figure}[H]
\centering
\begin{tabular}{cc}
\includegraphics[scale=0.42]{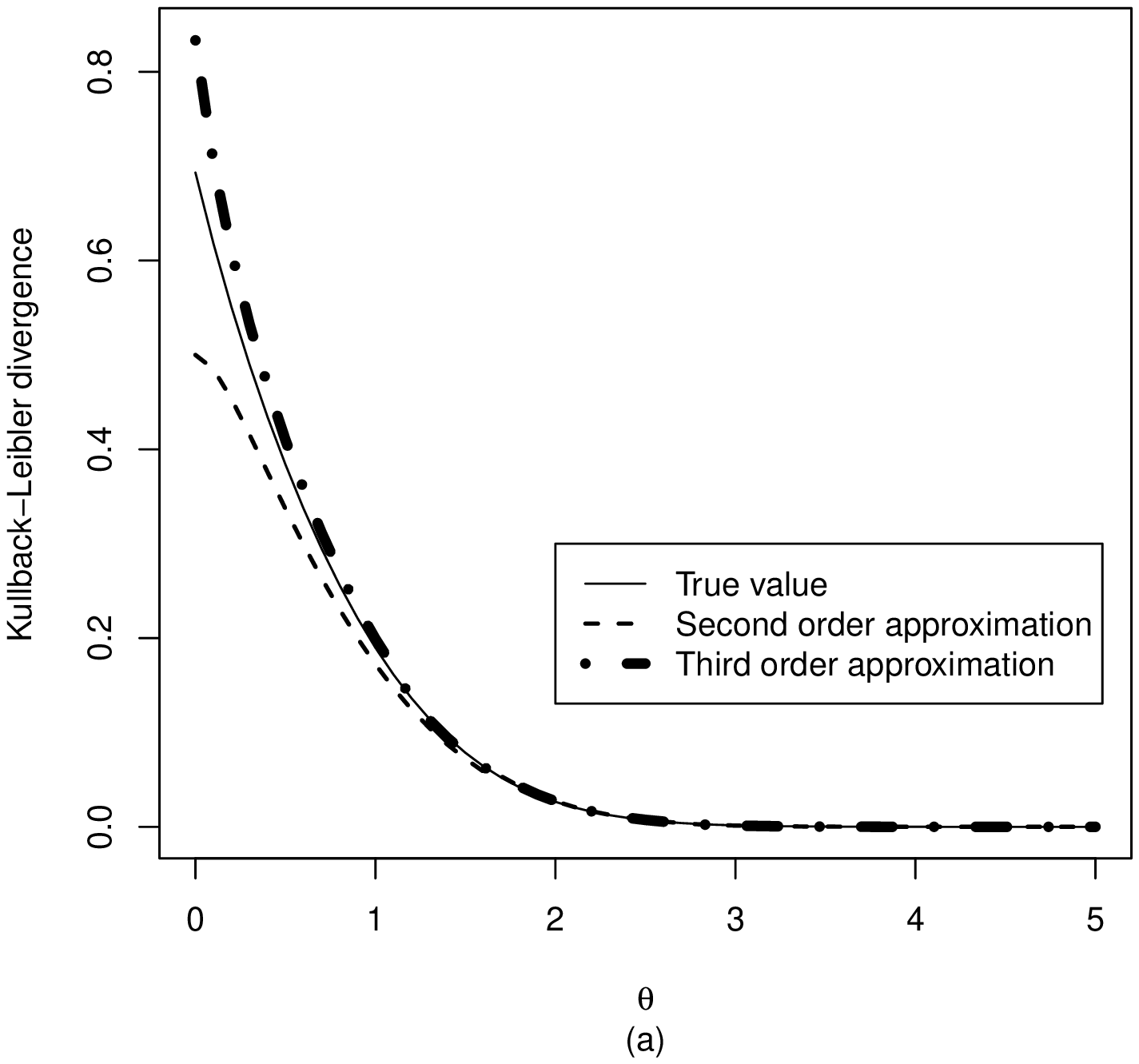} &
\includegraphics[scale=0.42]{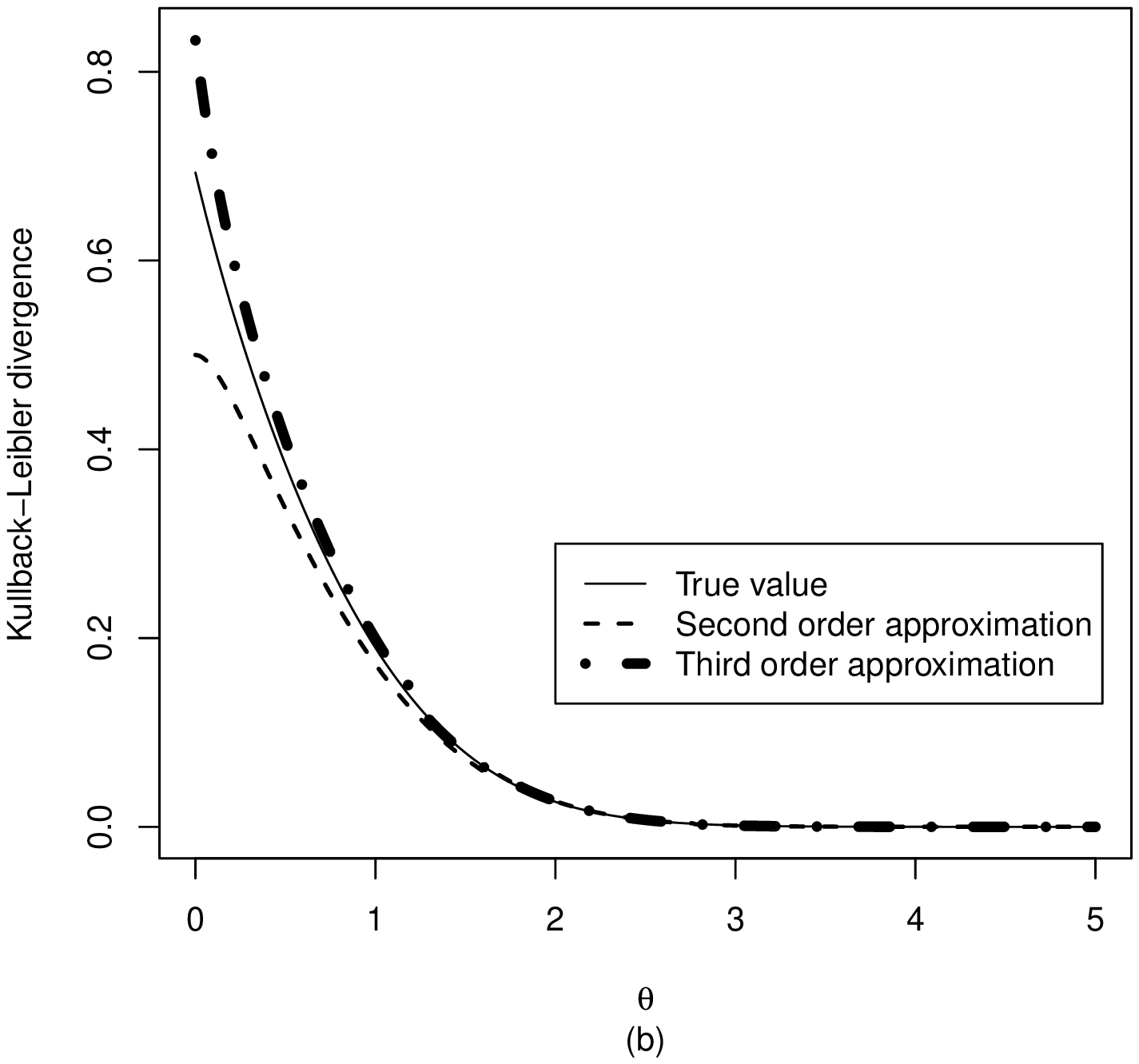}
\end{tabular}
\caption{Kullback--Leibler divergence from the normal for a range of values of $\theta=\frac{\mu}{\sigma}$ with $\sigma=1$ ({\bf a}) and $\sigma=5$ ({\bf b}).}
\label{kulgraph}
\end{figure}

Figure \ref{kulgraph} presents two cases of the Kullback--Leibler divergence, for illustration purposes, when the first two and three terms of the infinite sum have been used. In the first graph, the standard deviation is equal to one, and in the second case, it is equal to five. The divergence seems independent of the variance. The change occurs as a result of the value of $\theta$. It becomes clear that when the value of the mean to the standard deviation increases, the folded normal converges to the normal distribution.

\subsection{Kullback--Leibler Divergence from the Half Normal Distribution}

\begin{figure}[H]
\centering
\begin{tabular}{cc}
\includegraphics[scale=0.5]{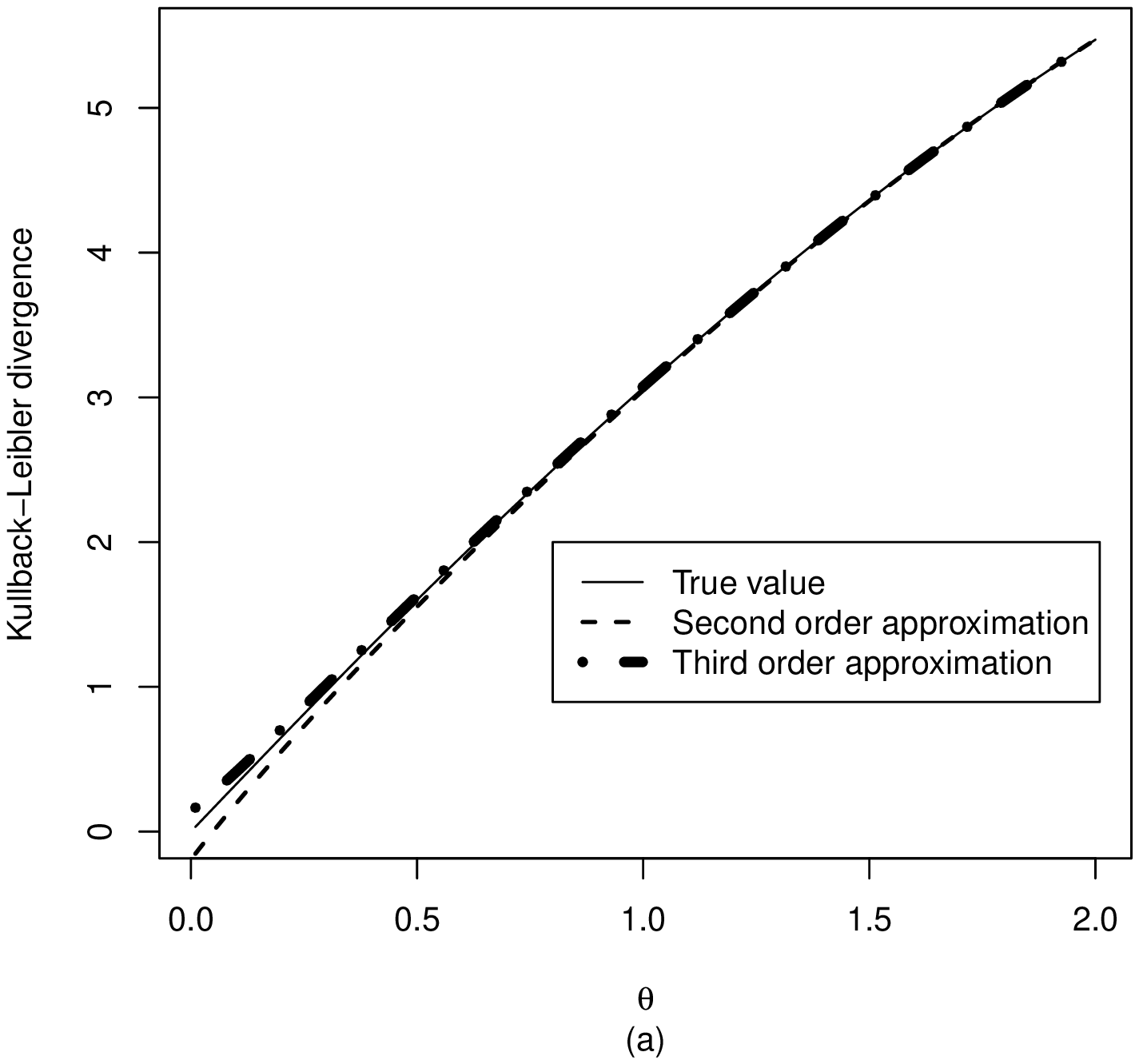} &
\includegraphics[scale=0.5]{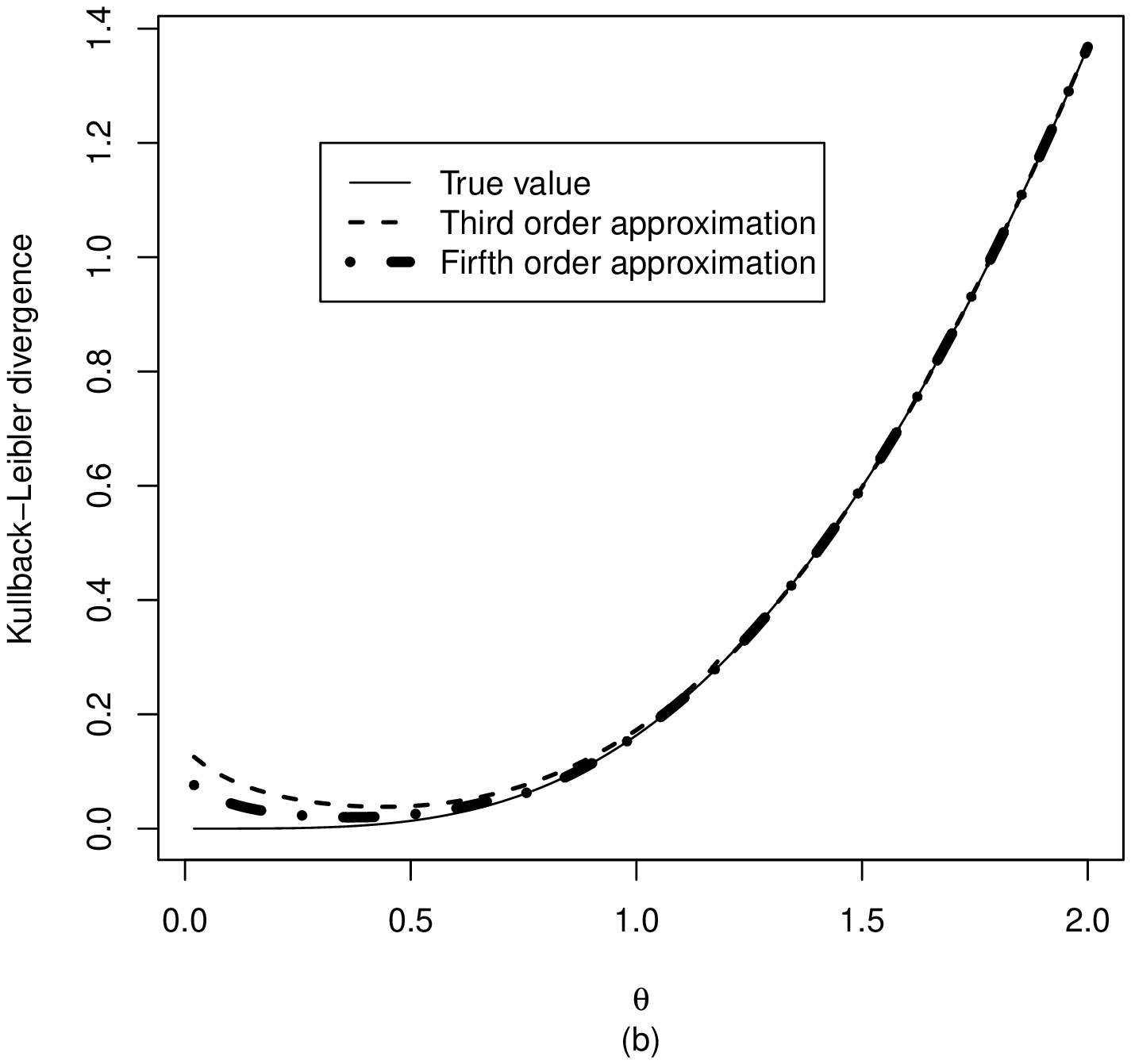}
\end{tabular}
\caption{Kullback--Leibler divergence from the half normal for a range of values of $\theta=\frac{\mu}{\sigma}$ with $\sigma=1$ ({\bf a}) and $\sigma=5$ ({\bf b}).}
\label{kulgraph2}
\end{figure}

As mentioned in Section \ref{relations}, the half normal distribution is a special case of the folded normal distribution with $\mu=0$. The Kullback--Leilber divergence of the folded normal from the half normal distribution is equal to:
\begin{eqnarray*}
& & KL(FN\left(\mu,\sigma^2\right)\vert \vert FN\left(\mu=0,\sigma^2\right))= \\
&= & \int_0^\infty\frac{1}{\sqrt{2\pi\sigma^2}}\left[ e^{-\frac{1}{2\sigma^2}\left(x-\mu \right)^2}+
e^{-\frac{1}{2\sigma^2}\left(x+\mu \right)^2} \right]\log{\frac{\frac{1}{\sqrt{2\pi\sigma^2}}\left[ e^{-\frac{1}{2\sigma^2}\left(x-\mu \right)^2}+
e^{-\frac{1}{2\sigma^2}\left(x+\mu \right)^2} \right]}{\frac{2}{\sqrt{2\pi\sigma^2}}e^{-\frac{1}{2\sigma^2}x^2}}} dx \\
&=& -\log{2} \int_0^\infty f\left(x;\mu,\sigma^2\right)dx+
\int_0^\infty f\left(x;\mu,\sigma^2\right)\log{\left(e^{-\frac{\mu^2}{2\sigma^2}+\frac{\mu x}{\sigma^2}}+
e^{-\frac{\mu^2}{2\sigma^2}-\frac{\mu x}{\sigma^2}}\right)} dx \\
&=& -\log{2}+\int_0^\infty \left(\frac{\mu x}{\sigma^2}-\frac{\mu^2}{2\sigma^2}\right)f\left(x;\mu,\sigma^2\right)dx+
\int_0^\infty f\left(x;\mu,\sigma^2\right)\log{\left(1+e^{-\frac{2\mu x}{\sigma^2}} \right)}dx \\
& = & -\log{2}+\frac{2\mu \mu_f-\mu^2}{2\sigma^2}+KL(FN \vert \vert N)
\end{eqnarray*}
where $f\left(x;\mu,\sigma^2\right)$ stands for the folded normal Equation (\ref{fd}) and $\mu_f$ is the expected value given in Equation (\ref{mean}).
Figure \ref{kulgraph2} shows the approximations to the true value when $\sigma=1$ and $\sigma=5$. This time, we used the third and fifth order approximations, but even then, for small values of $\theta$, the approximations were not satisfactory.

The previous result cannot lead to an inequality regarding the Kullback--Leibler divergences from the two other distributions. When $\mu>\sigma$, then the divergence from the half normal will be greater than the divergence from the normal, and when $\mu<\sigma$, the opposite is true. However, this is not strict, since it can be the case for either inequality that the relationship between the divergences is not true. Instead, we can use it as a rule of thumb in general.

\section{Parameter Estimation}

We will show two ways of estimating the parameters. The first one can be found in \cite{leone1961}, but we review it and add some more details. Both of them are essentially the maximum likelihood estimation procedure, but in the first case, we perform maximization, whereas in the second case, we seek the root of an equation.

The log-likelihood of Equation (\ref{fd}) can be written in the following way:
\begin{eqnarray} \label{loglik}
l &=& -\frac{n}{2}\log{2\pi\sigma^2}+\sum_{i=1}^n\log{\left[e^{-\frac{\left(x_i-\mu\right)^2}{2\sigma^2}}+
e^{-\frac{\left(x_i+\mu\right)^2}{2\sigma^2}} \right] } \Rightarrow \nonumber \\
l &=& -\frac{n}{2}\log{2\pi\sigma^2}+\sum_{i=1}^n\log{\left[e^{-\frac{\left(x_i-\mu\right)^2}{2\sigma^2}}
\left(1+e^{-\frac{\left(x_i+\mu\right)^2}{2\sigma^2}}e^{\frac{\left(x_i-\mu\right)^2}{2\sigma^2}}\right)\right]} \Rightarrow \nonumber \\
l &=& -\frac{n}{2}\log{2\pi\sigma^2}-\sum_{i=1}^n\frac{\left(x_i-\mu\right)^2}{2\sigma^2}+\sum_{i=1}^n\log{\left(1+e^{-\frac{2\mu x_i}{\sigma^2}} \right)}
\end{eqnarray}
where $n$ is the sample size of the $x_i$ values. The partial derivatives of Equation (\ref{loglik}) are:
\begin{eqnarray*}
\frac{\partial l}{\partial \mu} &=& \frac{\sum_{i=1}^n\left(x_i-\mu \right)}{\sigma^2}-
\frac{2}{\sigma^2}\sum_{i=1}^n\frac{x_ie^{\frac{-2\mu x_i}{\sigma^2}}}{1+e^{\frac{-2\mu x_i}{\sigma^2}}}=
\frac{\sum_{i=1}^n\left(x_i-\mu \right)}{\sigma^2}-\frac{2}{\sigma^2}\sum_{i=1}^n\frac{x_i}{1+e^{\frac{2\mu x_i}{\sigma^2}}}, \ \ \text{and} \\
\frac{\partial l}{\partial \sigma^2} &=& -\frac{n}{2\sigma^2}+\frac{\sum_{i=1}^n\left(x_i-\mu \right)^2}{2\sigma^4}+
\frac{2\mu}{\sigma^4}\sum_{i=1}^n\frac{x_ie^{-\frac{2\mu x_i}{\sigma^2}}}{1+e^{-\frac{2\mu x_i}{\sigma^2}}} \Rightarrow \\
\frac{\partial l}{\partial \sigma^2} &=& -\frac{n}{2\sigma^2}+\frac{\sum_{i=1}^n\left(x_i-\mu \right)^2}{2\sigma^4}+
\frac{2\mu}{\sigma^4}\sum_{i=1}^n\frac{x_i}{1+e^{\frac{2\mu x_i}{\sigma^2}}}
\end{eqnarray*}
By equating the first derivative of the log-likelihood to zero, we obtain a nice relationship:
\begin{eqnarray} \label{eq1}
\sum_{i=1}^n\frac{x_i}{1+e^{\frac{2\mu x_i}{\sigma^2}}}=\frac{\sum_{i=1}^n\left(x_i-\mu \right)}{2}
\end{eqnarray}
Note that Equation (\ref{eq1}) has three solutions, one at zero and two more with the opposite sign. The example in Section \ref{sim.ex} will show graphically the three solutions. By substituting Equation (\ref{eq1}), to the derivative of the log-likelihood w.r.t%please define
 $\sigma^2$ and equating to zero, we get the following expression for \linebreak the variance:
\begin{eqnarray} \label{eq2}
\sigma^2=\frac{\sum_{i=1}^n\left(x_i-\mu\right)^2}{n}+\frac{2\mu\sum_{i=1}^n\left(x_i-\mu\right)}{n}=\frac{\sum_{i=1}^n\left(x_i^2-\mu^2\right)}{n}=\frac{\sum_{i=1}^nx_i^2}{n}-\mu^2
\end{eqnarray}
The above relationships Equations (\ref{eq1}) and (\ref{eq2}) can be used to obtain maximum likelihood estimates in an efficient recursive way. We start with an initial value for $\sigma^2$ and find the positive root of \linebreak Equation (\ref{eq1}). Then, we insert this value of $\mu$ in Equation (\ref{eq2}) and get an updated value of $\sigma^2$. The procedure is being repeated until the change in the log-likelihood value is negligible.

Another easier and more efficient way is to perform a search algorithm. Let us write Equation (\ref{eq1}) in a more elegant way.
\begin{eqnarray*}
2\sum_{i=1}^n\frac{x_i}{1+e^{\frac{2\mu x_i}{\sigma^2}}}-
\sum_{i=1}^n\frac{x_i\left(1+e^{\frac{2\mu x_i}{\sigma^2}}\right)}{1+e^{\frac{2\mu x_i}{\sigma^2}}}+n\mu &=& 0 \Rightarrow \\
\sum_{i=1}^n\frac{x_i\left(1-e^{\frac{2\mu x_i}{\sigma^2}}\right)}{1+e^{\frac{2\mu x_i}{\sigma^2}}}+n\mu &=& 0
\end{eqnarray*}
where $\sigma^2$ is defined in Equation (\ref{eq2}).
It becomes clear that the optimization the log-likelihood \linebreak Equation (\ref{loglik}) with respect to the two parameters has turned into a root search of a function with one parameter only. We tried to perform maximization via the E-M %please define
algorithm, treating the sign as the missing information, but it did not prove very good in this case.

\subsection{An Example with Simulated Data} \label{sim.ex}

\begin{figure}[H]
\centering
\begin{tabular}{cc}

\includegraphics[scale=0.45]{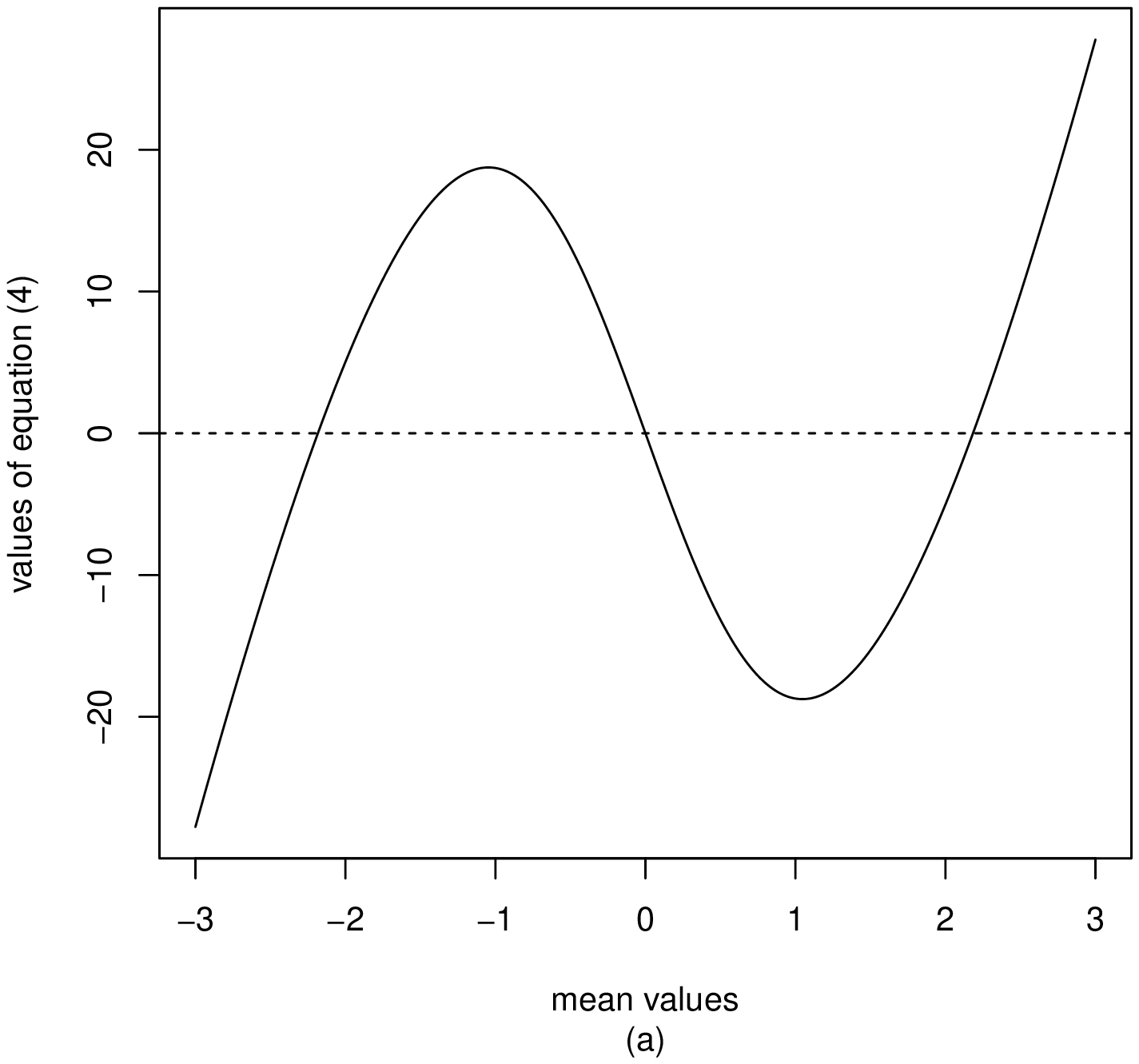} &
\includegraphics[scale=0.45]{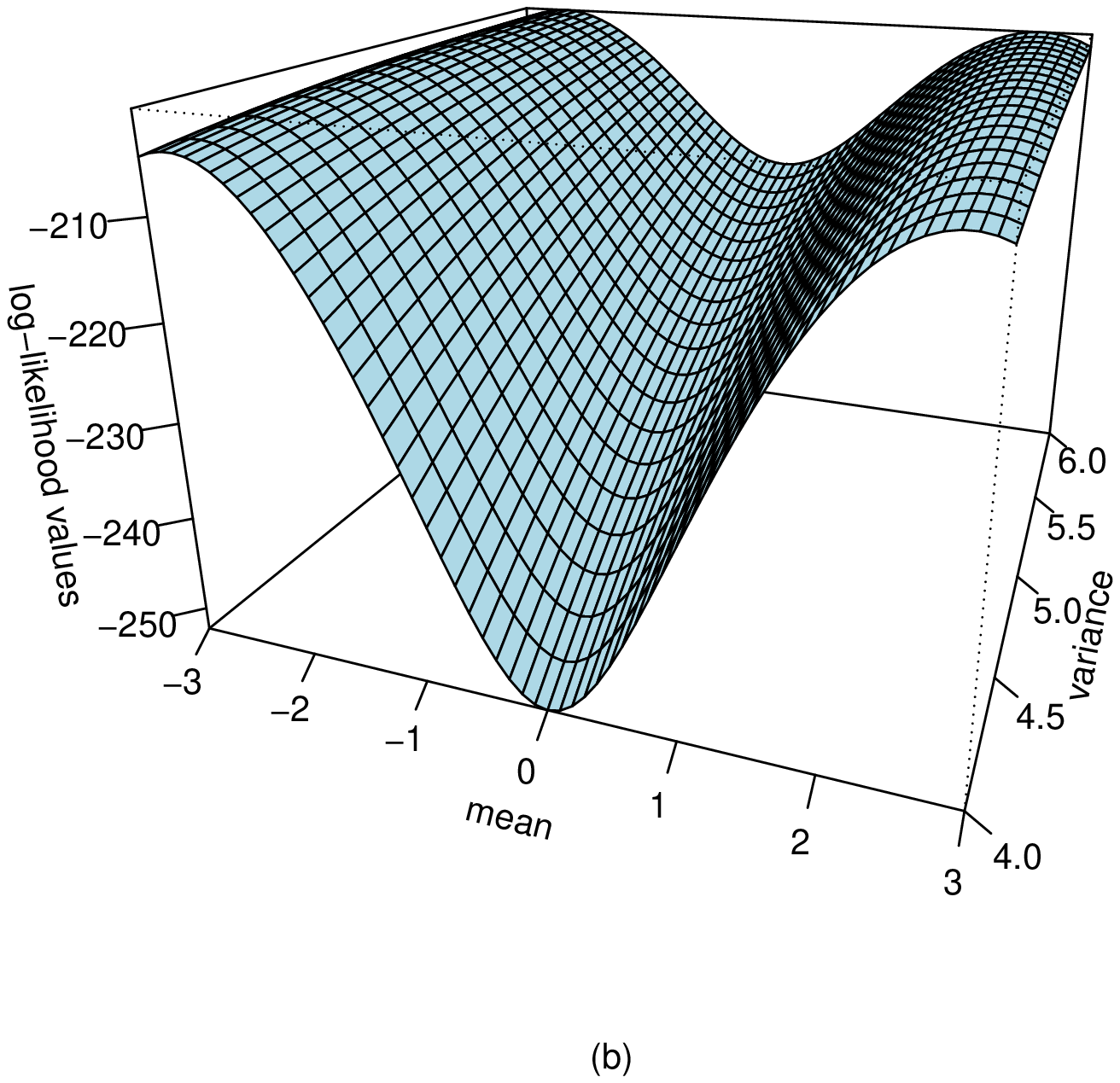}
\end{tabular}
\caption{The left graph ({\bf a})  shows the three solutions of the log-likelihood. The right three-dimensional figure ({\bf b})  shows the values of the log-likelihood for a range of mean and \protect\linebreak variance values.}
\label{ex_1}
\end{figure}

We generated $100$ random values from the $FN(2,9)$ in order to illustrate the maximum likelihood estimation procedure. The estimated parameter values were equal to $\left(\hat{\mu}=2.183, \hat{\sigma}^2=8.065\right)$. The corresponding $95\%$ confidence intervals for $\mu$ and $\sigma^2$ were $\left(0.782, 3.585\right)$ and $\left(2.022, 14.108\right)$ respectively. Figure \ref{ex_1} shows graphically the existence of the three extrema of the log-likelihood \linebreak Equation (\ref{loglik}), one minimum (always at zero) and two maxima at the maximum likelihood \linebreak estimates of $\mu$.

\subsection{Simulation Studies}

Simulation studies were implemented to examine the accuracy of the estimates using numerical optimization based on the simplex method \citep{nelder1965}. Numerical optimization was performed in \cite{R2012}, using the \textit{optim} function. The term accuracy refers to interval estimation rather than point estimation, since the interest was on constructing confidence intervals for the parameters. The number of simulations was set equal to R = 1,000. The sample sizes ranged from 20 to 100 for a range of values of the parameter vector. The R-package \textit{VGAM}%please define; are the italics necessary?
 \cite{yee2010} offers algorithms for obtaining maximum likelihood estimates of the folded normal, but we have not used it here.

For every simulation, we calculated $95\%$ confidence intervals using the normal approximation, where the variance was estimated from the inverse of the observed information matrix. The maximum likelihood estimates are asymptotically normal with variance equal to the inverse of the Fisher's information. The sample estimate of this information is given by the second derivative (Hessian matrix) of the log-likelihood with respect to the parameter. This is an asymptotic confidence interval.

Bootstrap confidence intervals were also calculated using the percentile method \cite{efron1993}. For every simulation, we produced the bootstrap distribution of the data with $B=1000$ bootstrap repetitions. Thus, we calculated the $2.5\%$ lower and upper quantiles for each of the parameters. In addition, we calculated the correlations for every pair of the parameters.

Tables 1 to 4 present the coverage of the $95\%$ confidence intervals for the two parameters at different pairs of sample size and mean. The rows correspond to the sample size, whereas the columns correspond to the ratio $\theta=\frac{\mu}{\sigma}$, with $\sigma=5$ fixed.

\begin{table}[H]
\centering\small
\caption{Estimated coverage probability of the $95\%$ confidence intervals for the mean parameter, $\mu$, using the observed information matrix.}
\label{mat1}

\begin{tabular}{@{}ccccccccc@{}} \toprule

 & & & {\bf Values} & {\bf of } & \boldmath$\theta$ & & & \\
{\bf Sample size} & {\bf 0.5 } & {\bf 1} & {\bf 1.5} & {\bf 2} & {\bf 2.5} & {\bf 3} & {\bf 3.5} & {\bf 4 } \\ \midrule

20 & 0.689 & 0.930 & 0.955 & 0.931 & 0.926 & 0.940 & 0.930 & 0.948 \\
30 & 0.679 & 0.921 & 0.949 & 0.943 & 0.925 & 0.926 & 0.941 & 0.915 \\
40 & 0.690 & 0.916 & 0.936 & 0.933 & 0.941 & 0.948 & 0.944 & 0.928 \\
50 & 0.718 & 0.944 & 0.955 & 0.938 & 0.933 & 0.948 & 0.946 & 0.946 \\
60 & 0.699 & 0.950 & 0.968 & 0.948 & 0.949 & 0.941 & 0.942 & 0.946 \\
70 & 0.721 & 0.931 & 0.956 & 0.939 & 0.939 & 0.939 & 0.949 & 0.945 \\
80 & 0.691 & 0.930 & 0.950 & 0.940 & 0.946 & 0.936 & 0.945 & 0.939 \\
90 & 0.720 & 0.932 & 0.960 & 0.949 & 0.949 & 0.939 & 0.954 & 0.944 \\
100 & 0.738 & 0.945 & 0.949 & 0.938 & 0.943 & 0.926 & 0.946 & 0.952 \\ \bottomrule

\end{tabular}

\end{table}

What can be seen from Tables \ref{mat1} and \ref{mat2} is that whist the sample size is important, the value of $\theta$, the mean to standard deviation ratio, is more important. As this ration increase the coverage probability increases, as well, and reaches the desired nominal $95\%$. This is also true for the bootstrap confidence intervals, but the coverage is in general higher and increases faster as the sample size increases in contrast to the asymptotic confidence interval. What is more is that when the value of $\theta$ is less than one, the bootstrap confidence interval is to be preferred. When the value of $\theta$ becomes equal to or more than one, then both the bootstrap and the asymptotic confidence intervals produce similar coverages.

The results regarding the variance are presented in Tables \ref{mat3} and \ref{mat4}. When the value of $\theta$ is small, both ways of obtaining confidence intervals for this parameter are rather conservative. The bootstrap intervals tend to perform better, but not up to the expectations. Even when the value of $\theta$ is large, if the sample sizes are not large enough, the nominal coverage of $95\%$ is not attained.

\begin{table}[H]
\centering
\small
\caption{Estimated coverage probability of the bootstrap $95\%$ confidence intervals for the mean parameter, $\mu$, using the percentile method.}
\label{mat2}

\begin{tabular}{@{}ccccccccc@{}} \toprule

 & & & {\bf Values} & {\bf of } & \boldmath$\theta$ & & & \\
{\bf Sample size} & {\bf 0.5 } & {\bf 1} & {\bf 1.5} & {\bf 2} & {\bf 2.5} & {\bf 3} & {\bf 3.5} & {\bf 4 } \\ \midrule

20 & 0.890 & 0.925 & 0.939 & 0.921 & 0.918 & 0.940 & 0.929 & 0.942 \\
30 & 0.894 & 0.931 & 0.933 & 0.943 & 0.926 & 0.922 & 0.942 & 0.910 \\
40 & 0.910 & 0.925 & 0.927 & 0.933 & 0.941 & 0.947 & 0.946 & 0.928 \\
50 & 0.914 & 0.943 & 0.942 & 0.934 & 0.934 & 0.945 & 0.946 & 0.943 \\
60 & 0.904 & 0.949 & 0.953 & 0.950 & 0.941 & 0.938 & 0.943 & 0.944 \\
70 & 0.893 & 0.934 & 0.943 & 0.936 & 0.937 & 0.938 & 0.949 & 0.939 \\
80 & 0.918 & 0.940 & 0.939 & 0.939 & 0.944 & 0.935 & 0.946 & 0.938 \\
90 & 0.920 & 0.934 & 0.952 & 0.948 & 0.946 & 0.939 & 0.951 & 0.947 \\
100 & 0.918 & 0.940 & 0.936 & 0.932 & 0.946 & 0.925 & 0.945 & 0.949 \\
\bottomrule

\end{tabular}

\end{table}

\vspace{-6pt}

\begin{table}[H]
\centering
\small
\caption{Estimated coverage probability of the $95\%$ confidence intervals for the variance parameter, $\sigma^2$, using the observed information matrix.}
\label{mat3}

\begin{tabular}{ccccccccc} \toprule

 & & & {\bf Values} & {\bf of } & \boldmath$\theta$ & & & \\
{\bf Sample size} & {\bf 0.5 } & {\bf 1} & {\bf 1.5} & {\bf 2} & {\bf 2.5} & {\bf 3} & {\bf 3.5} & {\bf 4 } \\ \midrule

20 & 0.649 & 0.765 & 0.854 & 0.853 & 0.876 & 0.870 & 0.862 & 0.885 \\
30 & 0.697 & 0.794 & 0.870 & 0.898 & 0.892 & 0.898 & 0.894 & 0.896 \\
40 & 0.723 & 0.849 & 0.893 & 0.914 & 0.919 & 0.913 & 0.909 & 0.902 \\
50 & 0.751 & 0.867 & 0.916 & 0.907 & 0.911 & 0.924 & 0.899 & 0.912 \\
60 & 0.745 & 0.865 & 0.911 & 0.913 & 0.916 & 0.906 & 0.920 & 0.933 \\
70 & 0.769 & 0.874 & 0.928 & 0.928 & 0.912 & 0.930 & 0.926 & 0.935 \\
80 & 0.776 & 0.883 & 0.927 & 0.919 & 0.934 & 0.936 & 0.916 & 0.924 \\
90 & 0.795 & 0.901 & 0.931 & 0.932 & 0.925 & 0.930 & 0.940 & 0.941 \\
100 & 0.824 & 0.904 & 0.927 & 0.933 & 0.925 & 0.936 & 0.932 & 0.942 \\ \bottomrule

\end{tabular}

\end{table}

The correlation between the two parameters was also estimated for every simulation from the observed information matrix. The results are displayed in Table \ref{matb}. The correlation between the two parameters is always negative irrespective of the sample size or the value of $\theta$, except for the case when $\theta=4$. In this case, the correlation becomes zero as expected. As the value of $\theta$ grows larger, the probability of the normal distribution, which lies on the negative axis, becomes smaller until it becomes negligible. In this case, the distribution equals the classical normal distribution for which the two parameters are known to be orthogonal.

\begin{table}[H]
\centering
\small
\caption{Estimated coverage probability of the bootstrap $95\%$ confidence intervals for the variance parameter, $\sigma^2$, using the percentile method.}
\label{mat4}

\begin{tabular}{ccccccccc} \toprule

 & & & {\bf Values} & {\bf of } & \boldmath$\theta$ & & & \\
{\bf Sample size} & {\bf 0.5 } & {\bf 1} & {\bf 1.5} & {\bf 2} & {\bf 2.5} & {\bf 3} & {\bf 3.5} & {\bf 4 } \\ \midrule

20 & 0.657 & 0.814 & 0.862 & 0.842 & 0.840 & 0.832 & 0.818 & 0.824 \\
30 & 0.701 & 0.850 & 0.885 & 0.891 & 0.882 & 0.867 & 0.869 & 0.866 \\
40 & 0.743 & 0.881 & 0.896 & 0.913 & 0.912 & 0.886 & 0.881 & 0.878 \\
50 & 0.772 & 0.895 & 0.921 & 0.916 & 0.897 & 0.901 & 0.885 & 0.892 \\
60 & 0.797 & 0.907 & 0.912 & 0.910 & 0.906 & 0.897 & 0.907 & 0.916 \\
70 & 0.807 & 0.904 & 0.925 & 0.915 & 0.909 & 0.918 & 0.908 & 0.924 \\
80 & 0.822 & 0.895 & 0.925 & 0.914 & 0.925 & 0.917 & 0.909 & 0.909 \\
90 & 0.869 & 0.916 & 0.932 & 0.922 & 0.919 & 0.915 & 0.934 & 0.929 \\
100 & 0.873 & 0.915 & 0.918 & 0.925 & 0.906 & 0.931 & 0.920 & 0.939 \\ \bottomrule

\end{tabular}

\end{table}

\begin{table}[H]
\centering
\small
\caption{Estimated correlations between the two parameters obtained from the observed information matrix.}
\label{matb}

\begin{tabular}{ccccccccc} \toprule

 & & & {\bf Values} & {\bf of } & \boldmath$\theta$ & & & \\
{\bf Sample size} & {\bf 0.5 } & {\bf 1} & {\bf 1.5} & {\bf 2} & {\bf 2.5} & {\bf 3} & {\bf 3.5} & {\bf 4 } \\ \midrule

20 &$-$0.600 &$-$0.495 &$-$0.272 &$-$0.086 &$-$0.025 &$-$0.006 &$-$0.001 & 0.000 \\
30 &$-$0.638 &$-$0.537 &$-$0.262 &$-$0.089 &$-$0.022 &$-$0.005 &$-$0.001 & 0.000 \\
40 &$-$0.695 &$-$0.548 &$-$0.251 &$-$0.081 &$-$0.021 &$-$0.005 &$-$0.001 & 0.000 \\
50 &$-$0.723 &$-$0.580 &$-$0.259 &$-$0.076 &$-$0.020 &$-$0.005 &$-$0.001 & 0.000 \\
60 &$-$0.750 &$-$0.597 &$-$0.251 &$-$0.075 &$-$0.019 &$-$0.004 &$-$0.001 & 0.000 \\
70 &$-$0.771 &$-$0.588 &$-$0.256 &$-$0.073 &$-$0.019 &$-$0.004 &$-$0.001 & 0.000 \\
80 &$-$0.774 &$-$0.604 &$-$0.253 &$-$0.074 &$-$0.019 &$-$0.004 &$-$0.001 & 0.000 \\
90 &$-$0.796 &$-$0.599 &$-$0.245 &$-$0.073 &$-$0.018 &$-$0.004 &$-$0.001 & 0.000 \\
100 &$-$0.804 &$-$0.611 &$-$0.252 &$-$0.072 &$-$0.019 &$-$0.004 &$-$0.001 & 0.000 \\ \bottomrule

\end{tabular}

\end{table}

Table \ref{probs} shows the probability of a normal random variable being less than zero when $\sigma=5$ and the same values of $\theta$ as in the simulation studies.

\begin{table}[H]
\centering
\small
\caption{Probability of a normal variable having negative values.}
\label{probs}

\begin{tabular}{cccccccc} \toprule

 & & {\bf Values} & {\bf of} & \boldmath$\theta$ & & & \\
{\bf 0.5 } & {\bf 1} & {\bf 1.5} & {\bf 2 } & {\bf 2.5} & {\bf 3 } & {\bf 3.5} & {\bf 4 } \\ \midrule

0.309 & 0.159 & 0.067 & 0.023 & 0.006 & 0.001 & 0.000 & 0.000 \\ \bottomrule

\end{tabular}

\end{table}

When the ratio of mean to standard deviation is small, the area of the normal distribution in the negative side is large, and as the value of this ratio increases, the probability decreases until it becomes zero. In this case, the folded normal is the normal distribution, since there are no negative values \linebreak to fold on to the positive side. This of course is in accordance with all the previous observations and results we saw.

\section{Application to Body Mass Index Data}

\begin{figure}[H]
\centering
\begin{tabular}{cc}
\includegraphics[scale=0.5]{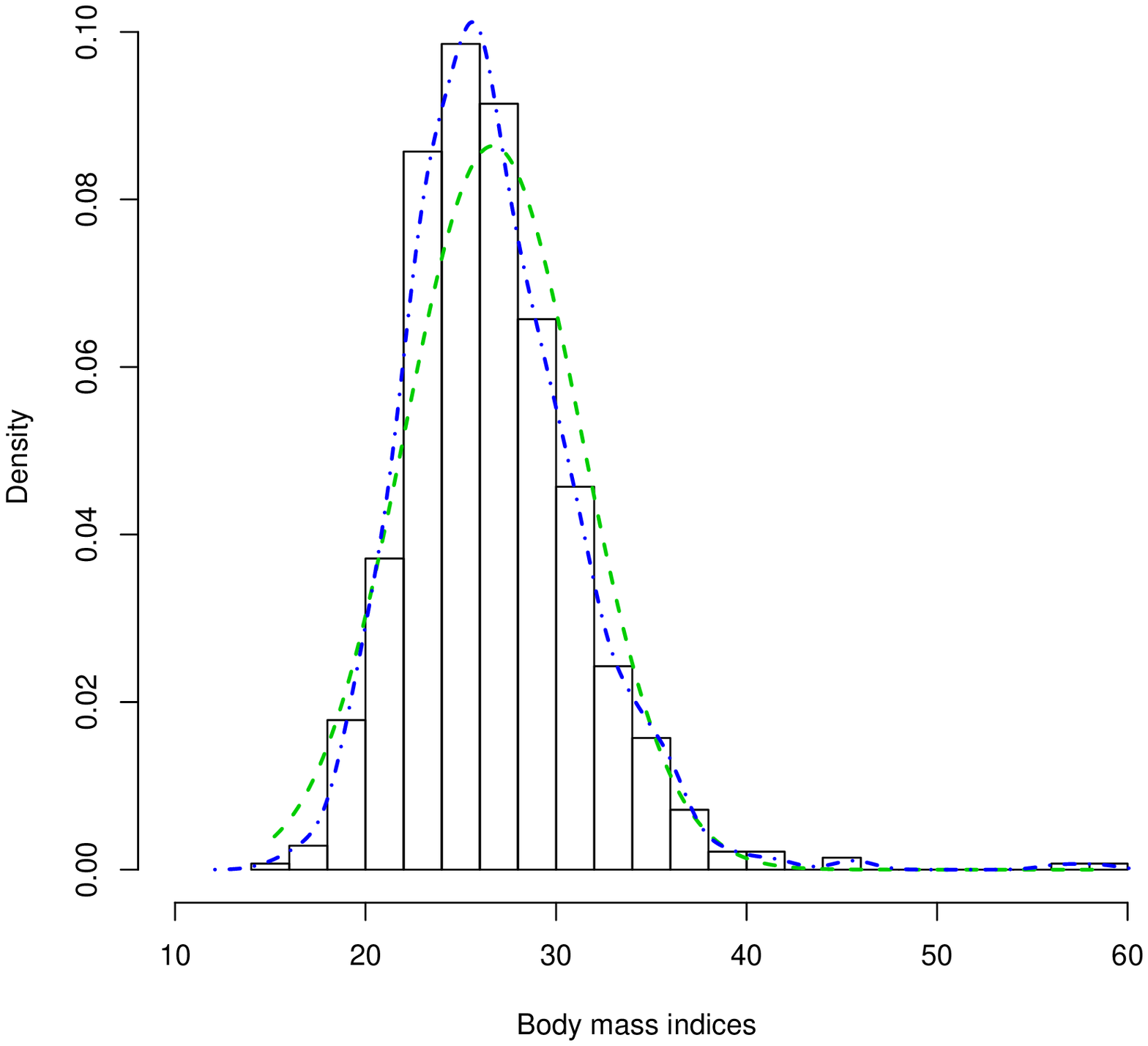} &
\includegraphics[scale=0.5]{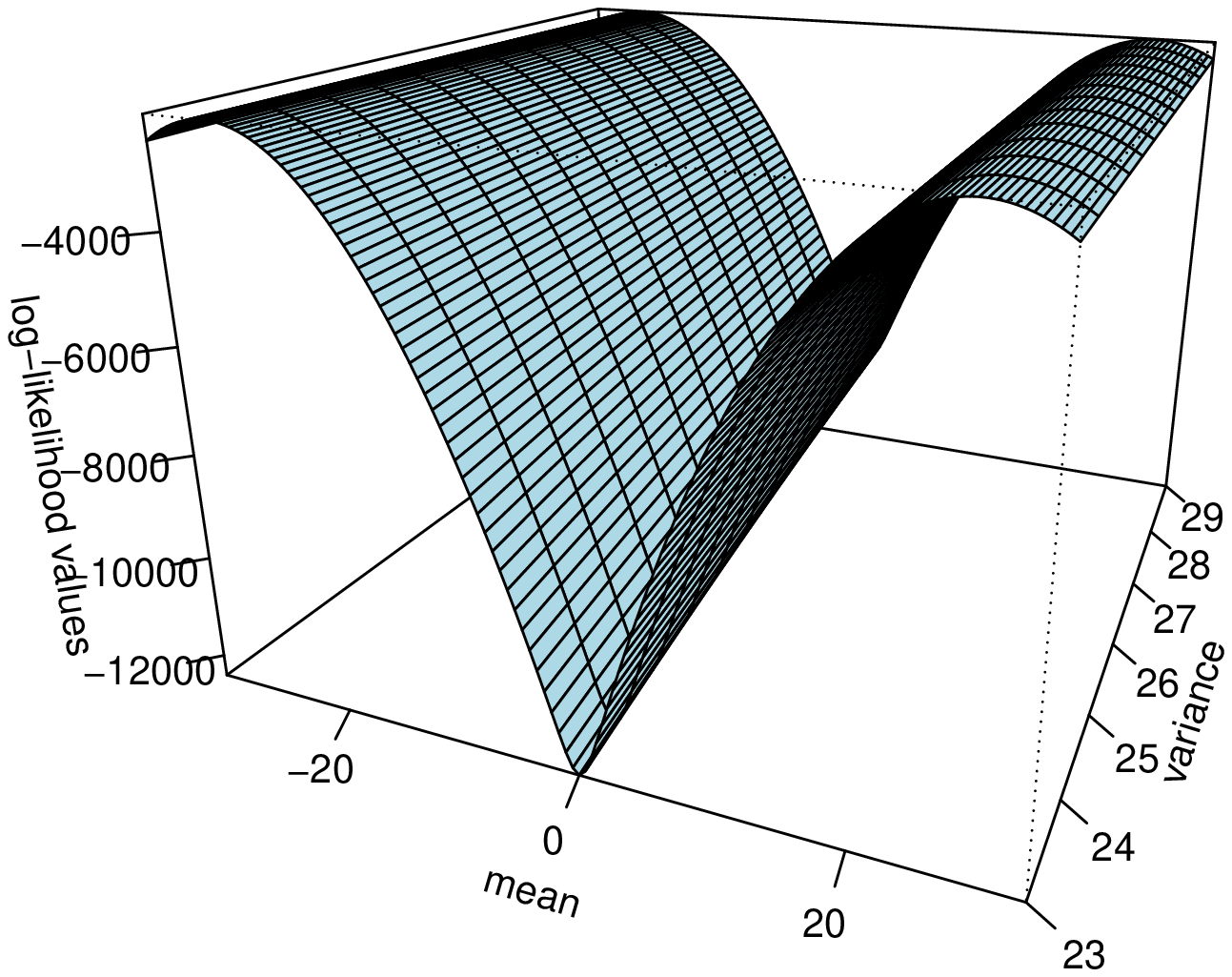}
\end{tabular}
\caption{The histogram on the left shows the body mass indices of $700$ New Zealand adults. The green line is the fitted folded normal and the blue line is the kernel density. The perspective plot on the right shows the log-likelihood of the body mass index data as a function of the mean and the variance.}
\label{ex_2}
\end{figure}

We fitted the folded normal distribution on real data. These are observations of the the
 body mass index of $700$ New Zealand adults, accessible via the R package \textit{VGAM} \citep{yee2010}.
 These measurements are a random sample from the Fletcher Challenge/Auckland Heart and Health survey conducted
 in the early 1990s \citep{macmahon1995}. Figure \ref{ex_2} contains a histogram of the data along
 with the parametric (folded normal) and the non-parametric (kernel) density estimation.
 It should be noted that the fitted folded normal
 here converges in distribution to the normal.

The estimated parameters (using the \textit{optim} command in R) were $\hat{\mu}=26.685 (0.175)$ and \linebreak $\hat{\sigma}^2=21.324
(1.140)$, with their standard error appearing inside the parentheses.
Since the sample size is very large, there is no need to estimate their standard errors and, consequently, $95\%$
confidence intervals, even though their ratio is only $1.251$. Their estimated correlation coefficient was very close to zero \scalebox{.95}[1.0]{($2$ $\times$ $10^{-4}$)}, and the estimated probability of the folded normal with these parameters below zero is \linebreak equal to zero.

\section{Discussion}

We derived the characteristic function of this distribution and, thus, its moment function. \linebreak The cumulant generating function is simply the logarithm of the moment generating function, and therefore, it is easy to calculate. The importance of these two functions is that they allow us to calculate all the moments of the distribution. In addition, we calculated the Laplace and Fourier transformations and the mean residual life.

The entropy of the folded normal distribution and the Kullback--Leibler divergence of this distribution from the normal and half normal distributions were approximated using the Taylor series. The results were numerically evaluated against the true values and were as expected.

We reviewed the maximum likelihood estimates and simplified their calculation and saw some properties of them. Confidence intervals for the parameters were obtained using the asymptotic theory and the bootstrap methodology under the umbrella of simulation studies.

The coverage of the confidence intervals for the two parameters was lower than the desired nominal in the small sample cases and when the mean to standard deviation ratio was lower than one. An alternative way to correct the under-coverage of the mean parameter is to use an alternative parametrization. The parameters $\theta=\frac{\mu}{\sigma}$ and $\sigma$ are calculated in \cite{johnson1962}. If we use $\theta$ and $\mu$, then the coverage of the interval estimation of $\mu$ is corrected, but the corresponding coverage of the confidence interval for $\sigma^2$ is still low.

The correlation between the two parameters was always negative and decreasing as the value of $\theta$ was increasing, as expected, until the two parameters become independent.

An application of the folded normal distribution to real data was exhibited, providing evidence that it can be used to model non-negative data adequately.

\section*{\noindent Conflicts of Interest}
\vspace{12pt}

The authors declare no conflict of interest.

\bibliographystyle{mdpi}

\begin{thebibliography}{9999}

%\begin{thebibliography}{9}

\bibitem{leone1961} Leone, F.C.; Nelson, L.S.; Nottingham, R.B. The folded normal distribution. {\it Technometrics} {\bf 1961}, {\it 3}, {543--550}.
\bibitem{Lin} Lin, H.C. The measurement of a process capability for folded normal process
data. {\it Int. J. Adv. Manuf. Technol.} {\bf 2004}, {\it 24}, 223--228.
\bibitem{Ashis} Chakraborty, A.K.; Chatterjee, M. On multivariate folded normal distribution.
 {\it Sankhya} {\bf 2013}, {\it 75}, 1--15.
\bibitem{elandt1961}Elandt, R.C. The folded normal distribution: Two methods of estimating parameters from moments. {\it Technometrics} {\bf 1961}, {\it 3}, 551--562.
\bibitem{johnson1962} Johnson, N.L. The folded normal distribution: Accuracy of estimation by maximum likelihood. {\it Technometrics} {\bf 1962}, {\it 4}, 249--256.
\bibitem{johnson1963} Johnson, N.L. Cumulative sum control charts for the folded normal distribution. {\it Technometrics} {\bf 1963}, {\it 5}, 451--458.
\bibitem{sundberg1974} Sundberg, R. On estimation and testing for the folded normal distribution. {\it Commun. Stat.-Theory Methods} {\bf 1974}, {\it 3}, 55--72.

\bibitem{KK} Kim, H.J. On the ratio of two folded normal distributions. {\it Commun. Stat.-Theory Methods} {\bf 2006}, {\it 35}, 965--977.


\bibitem{liao2010} Liao, M.Y. Economic tolerance design for folded normal data. {\it Int. J. Prod. Res.} {\bf 2010}, {\it 48}, 4123--4137.

\bibitem{nelder1965} Nelder, J.A.; Mead, R. A simplex method for function minimization. {\it Comput. J.} {\bf 1965}, {\it 7}, \linebreak 308--313.

\bibitem{johnson1994} Johnson, N.L; Kotz, S.; Balakrishnan, N. {\it Continuous Univariate Distributions}; John Wiley \& Sons, Inc.: New York, NY, USA, 1994.

\bibitem{psarakis1990} Psarakis, S.; Panaretos, J. The folded t distribution. {\it Commun. Stat.-Theory Methods} {\bf 1990}, {\it 19}, 2717--2734.

\bibitem{psarakis2000} Psarakis, S.; Panaretos, J. On some bivariate extensions of the folded normal and the folded t distributions. {\it J. Appl. Stat. Sci.} {\bf 2000}, {\it 10}, 119--136.

\bibitem{Kullback1997} Kullback, S. {\it Information Theory and Statistics}; Dover Publications: New York, NY, USA, 1977.



%%%%%%%%%%%%%%%%%%%%%%%%%%%%%%

\bibitem{R2012} R Development Core Team. R: A Language and Environment for Statistical Computing, 2012. Available online: http://www.R-project.org/ (accessed on 1 December 2013).



\bibitem{yee2010} Yee, T.W. The VGAM package for categorical data analysis. {\it J. Stat. Softw.} {\bf 2010}, {\it 32}, 1--34.









\bibitem{efron1993} Efron, B.; Tibshirani, R. {\it An Introduction to the Bootstrap}; Chapman and Hall/CRC: New York, NY, USA, 1993.
\bibitem{macmahon1995} MacMahon, S.; Norton, R.; Jackson, R.; Mackie, M.J.; Cheng, A.; Vander Hoorn, S.; \linebreak Milne, A.; McCulloch, A. Fletcher challenge-university of Auckland heart and health study: Design and baseline findings. {\it N. Zeal. Med. J.} {\bf 1995}, {\it 108}, 499--502.


\end{thebibliography}
\makeatletter
\renewcommand\@biblabel[1]{#1. }
\makeatother

\end{document}